\input amssym.def
\input amssym
\input epsf

\input harvmac.tex
\def\half{{1\over2}}
\def\ep{\tilde{\epsilon}}
\def\ol{\tilde{L}}
\def\t{\theta}

\def\vpint{{\rm -}\kern -1.1em\int_{-\infty}^\infty}
\def\happa{\chi}
\def\fanm{\phantom{-\,}}\def\gap{$\hfill$}
\def\capt#1{\narrower{\baselineskip=14pt plus 1pt minus 1pt\footnotefont #1}}

\lref\FaTa{ Faddeev, L.D. and  Takhtajan, L.A.: Hamiltonian Method in the
Theory of Solitons. New York: Springer 1987}

\lref\ZaZa{Zamolodchikov, A.B. and Zamolodchikov, Al.B.:
Factorized S-matrices in two dimensions as the exact
solutions of certain relativistic quantum field theory models.
Ann. Phys. (N.Y.) {\bf 120}, 253-291 (1979) }

\lref\ReSm{Reshetikhin, N.Yu. and Smirnov, F.A.:
Hidden quantum group symmetry and
integrable perturbations of conformal
field theories. Commun. Math. Phys.
{\bf 131}, 157-178 (1990)}

\lref\BeLeC{Bernard, D. and LeClair, A.:
Residual quantum symmetries of
the restricted Sine-Gordon theories.
Nucl. Phys. {\bf B340},
721-751 (1990)}

\lref\Rav{Fioravanti, D., Ravanini, F. and Stanishkov, M.:
Generalized KdV and Quantum Inverse Scattering Description
of Conformal Minimal Models. Preprint DFUB 95-08,
\#hep-th/9510047}

\lref\Sal{Fendley, P. and Saleur, H.:
Massless integrable quantum field theories 
and massless scattering in 1+1 dimensions.  Preprint USC-93-022, 
\#hep-th/9310058} 

\lref\jap{Sasaki, R. and Yamanaka, I.:
Virasoro algebra, vertex operators, quantum Sine-Gordon
and solvable Quantum Field theories. Adv. Stud. in Pure Math.
{\bf 16}, 271-296 (1988) }
\lref\Eguchi{Eguchi, T. and Yang, S.K.:
Deformation of conformal field theories and soliton
equations. Phys. Lett. {\bf B224}, 373-378  (1989)}
\lref\Frenk{Feigin, B. and Frenkel, E.:
Integrals of motion and quantum groups.
Proceeding of C.I.M.E. Summer School on
``Integrable systems and Quantum groups", \#hep-th/9310022  }

\lref\Kirillov{Kirillov, A.N.:
Dilogarithm identities. Progr. Theor. Phys. Supp.
{\bf 118}, 61-142 (1995)}

\lref\Confor{Zamolodchikov Al. B.: Two-Point Correlation Function in 
Scaling Lee-Yang Model. Nucl. Phys. {\bf B348}, 619-641  (1991)}

\lref\Bern{Abramowitz, M. and Stegun, I.:
Handbook of mathematical function, New York:
Dover publications, Inc. 1970 }

\lref\Cardyu{Cardy,J.L.:
Conformal invariance and the Yang-Lee edge
singularity in
two-dimensions.
Phys. Rev. Lett. {\bf 54}, 1354-1356 (1985)} 

\lref\Baxn{Baxter, R.J.:Partition function of the eight vertex model.
Ann. Phys. {\bf 70}, 193-228 (1972)}

\lref\Baxter{Baxter, R.J.: Exactly Solved Models
in Statistical Mechanics. London: Academic Press 1982}

\lref\Klassen{Klassen, T.R. and Melzer, E.:
Spectral flow between conformal field theories in\ $1+1$\
dimensions. Nucl. Phys. {\bf B370}, 511-570 (1992)}

\lref\KlassM{Klassen, T.R. and Melzer, E.:
On the Relation Between Scattering Amplitudes and
Finite Size Mass Corrections. Nucl. Phys. {\bf B362}, 329-388 (1991)}

\lref\BLZ{Bazhanov, V.V., Lukyanov, S.L. and Zamolodchikov, A.B.:
Integrable structure of conformal field theory,
quantum KdV theory and
thermodynamic Bethe ansatz.
Commun. Math. Phys. {\bf 177}, 381-398 (1996)}

\lref\Factor{ Mussardo, G.: Off-critical statistical models,
factorized scattering theories and bootstrap program.
Phys. Rep. {\bf 218}, 215-379 (1992)}

\lref\YY{Yang, C.N. and Yang, C.P.: Thermodynamics of one-dimensional
system of bosons with repulsive delta-function potential.
J. Math. Phys. {\bf 10}, 1115-1123 (1969)}

\lref\Zar{Zamolodchikov, Al.B.: Thermodynamic Bethe ansatz in
relativistic models: Scaling 3-state Potts and Lee-Yang models.
Nucl. Phys. {\bf B342}, 695-720 (1990)}

\lref\DotsFat{ Dotsenko, Vl.S. and Fateev, V.A.:
Conformal algebra and
multipoint correlation functions in 2d statistical models. Nucl. Phys.
{\bf B240}\ [{\bf FS12}], 312-348 (1984)\semi
Dotsenko, Vl.S. and  Fateev, V.A.: Four-point
correlation functions and the operator algebra in 2d conformal invariant
theories with central charge $c\le1$.
Nucl. Phys. {\bf B251}\ [{\bf FS13}] 691-734 (1985) }

\lref\Fei{Feigin, B.L. and  Fuchs, D.B.  Representations
of the Virasoro algebra. In:
Faddeev, L.D., Mal'cev, A.A. (eds.) Topology.
Proceedings, Leningrad 1982.
Lect. Notes in Math.
{\bf 1060}.
Berlin, Heidelberg, New York: Springer 1984}

\lref\ABZ{Zamolodchikov, A.B.:
Integrable field theory from conformal field theory.
Adv. Stud. in Pure Math.
{\bf 19}, 641-674 (1989)}

\lref\SFN{Fendley, P. and Saleur, H.:
Exact perturbative solution of the Kondo problem.
Phys. Rev. Lett. {\bf 75},  4492-4495 (1995) }

\lref\FLS{Fendley, P.,   Ludwig, A.W.W. and Saleur, H.:
Exact non-equilibrium 
transport through point contact 
in quantum wires and fractional Hall devices.
preprint USC-95/007, \#cond-mat/9503172.}

\lref\Zams{ Zamolodchikov, Al.B.:
On the TBA Equations for Reflectionless ADE
Scattering Theories.
Phys. Lett. {\bf B253}, 391 (1991)}

\lref\Yur{Yurov, V.P. and Zamolodchikov, Al.B.:
Truncated conformal space approach to scaling Lee-Yang model.
Int. J. Mod. Phys. {\bf A5}, 3221-3245 (1990)}

\lref\BaxPea{Baxter, R.J., Pearce, P.A.:
Hard Hexagons: Interfacial tension and correlation length.
J. Phys. A: Math. Gen. {\bf 15}   897-910 (1982)}

\lref\KluP{Kl\"umper, A. and Pearce, P.A.:
Analytical calculations of Scaling Dimensions: Tricritical Hard Square 
and Critical Hard Hexagons. J. Stat. Phys. {\bf 64}, 13-76 (1991)\hfill
\break
Kl\"umper, A. and Pearce, P.A.:
Conformal weights of RSOS lattice models and their
fusion hierarchies. J. Phys. {\bf A183}, 304-350  (1992) }

\lref\BLZZ{Bazhanov, V.V., Lukyanov, S.L. and Zamolodchikov, A.B.:
Integrable Structure of Conformal Field Theory II.
Q-operator and DDV equation.
Preprint CLNS 96/1405, LPTENS 96/18,
\#hepth  9604044}

\lref\Fendly{Fendley, P.: Exited state thermodynamics.
Nucl. Phys. {\bf B374}, 667-691 (1992)}

\lref\Martins{ Martins, M.J.:
Complex excitations in the Thermodyanamic Bethe Ansatz
approach. 
Phys. Rev. Lett. {\bf 67},   419-421 (1991)}

\lref\Musardo{Mussardo, G. and
Cardy, L.: S-matrix of the Yang-Lee edge singularity in
two-dimensions. Phys. Lett. {\bf B225}, 275-278, 1989}

\lref\Zarn{Zamolodchikov, Al.B.: Mass scale
in the Sine-Gordon model and its
reductions. Int. J. Mod. Phys. {\bf A10}, 1125-1150  (1995) }

\lref\BPZ{Belavin, A.A., Polyakov, A.M. and
Zamolodchikov, A.B.:
Infinite conformal symmetry in two-dimensional
quantum field theory.
Nucl. Phys. {\bf B241}, 333-380 (1984)}

\lref\BLZZZ{Bazhanov, V.V., Lukyanov, S.L. and Zamolodchikov, A.B.:
Integrable Structure of Conformal Field Theory III. The Yang-Baxter
Relations. To appear.}

\lref\BLZZZZ{Bazhanov, V.V., Lukyanov, S.L.
and Zamolodchikov, A.B.: Work in
progress.}

\Title{\vbox{\baselineskip12pt\hbox{CLNS 96/1416 }
\hbox{LPM-96/24}
\hbox{hep-th/9607099}}}
{\vbox{\centerline
{Integrable Quantum Field Theories  in Finite Volume:}
\vskip6pt\centerline{
Excited State Energies.}}}
\centerline{Vladimir V. Bazhanov$^1$\footnote{$^*$}
{e-mail address: Vladimir.Bazhanov@anu.edu.au},
Sergei L. Lukyanov$^{2,4}$ \footnote{$^{**}$}
{e-mail address: sergei@hepth.cornell.edu}}
\centerline{
and Alexander B. Zamolodchikov$^{3,4}$ \footnote{$^\dagger$}
{On leave of absence from Department of Physics and Astronomy,
Rutgers University, Piscataway,
NJ 08855-0849, USA} \footnote{$^\&$}{Guggenheim
Fellow} }
\centerline{$^1 $ Department of
Theoretical Physics and Center of Mathematics}
\centerline{and its Applications, IAS, Australian National University, }
\centerline{Canberra, ACT 0200, Australia}
\centerline{and}
\centerline{Saint Petersburg Branch of Steklov Mathematical Institute,}
\centerline{Fontanka 27, Saint Petersburg, 191011, Russia}
\centerline{$^2$Newman Laboratory, Cornell University}
\centerline{ Ithaca, NY 14853-5001, USA}
\centerline{$^3$Laboratoire de Physique
Math\'ematique, Universit\'e de Montpellier II,}
\centerline{ Pl. E. Bataillon,  34095 Montpellier, France}
\centerline{and}
\centerline{$^4$L.D. Landau Institute for Theoretical Physics,}
\centerline{Chernogolovka, 142432, Russia}

\Date{July, 96}
\eject

\centerline{{\bf Abstract}}
We develop a method of computing
the excited state energies in
Integrable Quantum Field Theories 
(IQFT) in finite geometry, with spatial 
coordinate compactified on
a circle of circumference $R$. The IQFT 
``commuting transfer-matrices''
introduced in\ \BLZ\ for  Conformal Field 
Theories (CFT) are generalized
to non-conformal IQFT obtained by perturbing
CFT with the operator $\Phi_{1,3}$.
 We study the models in which the 
fusion relations for these ``transfer-matrices''
truncate and provide
closed integral equations 
which generalize the equations of
Thermodynamic Bethe 
Ansatz to excited states. The explicit calculations 
are done for the first 
excited state in the ``Scaling Lee-Yang Model''.

\vfill
\eject
\newsec{Introduction}

Much progress in understanding $2D$
Integrable Quantum Field Theories 
(IQFT) has been achieved during the last two decades
(see e.g. \Factor\ for a review). In particular, many
models have been exactly solved ``on-shell" i.e. their mass spectra and 
their factorizable $S$-matrices have been determined. Important progress 
was also made towards ``off-shell" solution of IQFT. The approach based 
on so called Thermodynamic Bethe Ansatz (TBA) technique
\ \YY\  has proven to 
be very successful \Zar . This approach allows one to calculate the ground
state energy $E_0 (R)$ of
IQFT in the finite-size geometry, where the
spatial coordinate is compactified on the circle of circumference $R$,
provided the ``on-shell" solution is known. The problem being reduced
to solving certain non-linear integral equation (TBA equation). The 
quantity $E_0 (R)$ makes it possible to probe the scale dependence
in the IQFT, in particular, to find in the limit $R\to 0$ the central
charge $c$ of the Conformal Field Theory (CFT) which governs the short
distance behavior of the IQFT.

So far the power of TBA approach was limited to the ground state energy
$E_0 (R)$ (or the ground state energies in
``twisted'' sectors \Martins, \Fendly ). At the
same time it seems to be interesting to calculate the full energy 
spectrum $E_n (R)$ of the finite-size IQFT. These energies would 
provide interpolation between the CFT spectrum (determined by the CFT 
central charge $c$ and the conformal
dimensions $\Delta_a$) at $R\to 0$ and 
massive spectrum (determined by the particle masses and the $S$-matrix
amplitudes) at $R\to \infty$. However it is not clear how the ideas
behind traditional derivation of TBA equation could be generalized to
incorporate the excited states.

In our recent paper\ \BLZ\  
we have shown how one can define the family of
operators ${\bf T}_j (\lambda);\ j=1/2,1,3/2,\ldots$ --- the ``QFT 
transfer-matrices" \foot
{These operators appear as the continuous QFT versions of Baxter's 
commuting transfer-matrices
\ \Baxn,\  \Baxter\ 
and therefore we maintain using this term 
although the original meaning of the term ``transfer-matrix" here 
is apparently lost.} which act in the space of states of the finite size
IQFT and commute for different values of the parameter $\lambda$
\eqn\Tcomm{\big[{\bf T}_j (\lambda),
{\bf T}_{j'} (\lambda')\big] =0\ .}
We have also shown that the ``fusion relations" for the operators
${\bf T}_j (\lambda)$
\eqn\Tfusion{
{\bf T}_j (q^{1\over 2}\lambda) {\bf T}_j (q^{-{1\over 2}}\lambda)
= {\bf I} + {\bf T}_{j-1/2}(\lambda){\bf T}_{j+1/2}(\lambda)\ .}
(here $q$ is a parameter related to the Virasoro central charge $c$),
give rise to the functional equations for the eigenvalues of these 
``transfer-matrices" and that for the case
of the ground state eigenvalue these  reduce to the TBA
equations; at the same time the functional equations apply to the
excited states as well and following
the method developed by Kl\"umper
and Pearce\ \KluP\  
one can derive from them the nonlinear integral equations
which determine in principle the excited states $E_n (R)$. 

Our analysis in\ \BLZ\  explicitly concerns the case of CFT (more
precisely, the $c<1$ ``minimal" CFT\ \BPZ), where the space of states is 
decomposed in the sum of the direct products of ``right" and ``left" 
irreducible highest-weight Virasoro module ${\cal V}_{\Delta_a}$ and  
the operators ${\bf T}(\lambda)$ act in each of these spaces 
${\cal V}_{\Delta_a}$ separately. However, one can consider more
general non-conformal IQFT which are obtained by perturbing the 
``minimal CFT'' with the operator $\Phi_{1,3}$ \ABZ. In this paper
we show that the commuting operators similar to ${\bf T}_j (\lambda)$
can be constructed in the perturbed theory as well (we denote these
``perturbed'' transfer matrices as ${\Bbb T}_j (\mu|\lambda)$, where $\mu$
is the parameter of the perturbation), and that the most important
properties of the operators ${\bf T}_j (\lambda)$
obtained in \BLZ,
including the fusion relation \Tfusion, remain valid for the 
``perturbed'' operators ${\Bbb T}_j (\mu|\lambda)$. Although the
analytic properties of the operators ${\Bbb T}_j (\mu|\lambda)$ as the
functions of $\lambda$ turn out to be significantly different from those
of the ``conformal'' operators ${\bf T}_j (\lambda)$, it is still
possible to derive the nonlinear integral equations which determine the
finite-size energy levels $E_n (R)$ in the perturbed theory.

To demonstrate the efficiency of our approach we calculate the 
energy $E_1 (R)$ of the first excited state of the so called 
Scaling Lee-Yang Model (SLYM). This model is obtained by perturbing
the $c=-22/5$ ``minimal'' CFT ${\cal M}_{2/5}$ with the operator
$\Phi = \Phi_{1,3}$ of the dimension $\Delta = -1/5$, i.e. it is 
defined by the action 
\eqn\SLYM{{\cal A}_{SLYM} = {\cal A}_{{\cal M}_{2/5}}   +{\hat \mu}^2\, 
\int \Phi (x)\,d^2 x\ ,}
where ${\cal A}_{{\cal M}_{2/5}}$ is the action of the CFT 
${\cal M}_{2/5}$ and 
${\hat \mu}$ is the coupling parameter. The QFT defined
by \SLYM\  is integrable \ \ABZ\
and its on-shell solution (i.e. the particle
content and the factorizable S-matrix) was found in \Musardo.
The theory
contains one sort of scalar particles with the mass $m$ and their 
two-particle S-matrix is
\eqn\Smatrix{S(\theta) =
{{\sinh(\theta) + i\sin(\pi/3)}\over {\sinh(\theta) -
i\sin(\pi/3)}}\ ,}
where $\theta = \theta_1 - \theta_2$; $\theta_1$ and $\theta_2$ are
the rapidities of the scattering particles (i.e. their momenta are
$p_1 = m\,\sinh(\theta_1),\ p_2 = m\,\sinh(\theta_2)\, )$. The finite 
size ground state energy $E_0 (R)$ of  \SLYM\
was obtained in\ \Zar\  using 
TBA technique. Also, few first excited levels $E_n (R)$ was studied 
on the basis of ``Truncated Conformal Space method'' in \Yur. More
recently the exact relation between the mass $m$ and the coupling 
parameter ${\hat \mu}$ in \SLYM\  was found in \Zarn
\eqn\mumly{
\hat\mu^2=i\,\,{2^{1/5}\,5^{3/4}\over 16 \,\pi^{6/5}}\,\,
{\big(\Gamma (2/3)\Gamma (5/6)\big)^{12/5}
\over\Gamma (3/5)\Gamma (4/5)}\,\,m^{12/5}\ .
}
We apply the approach based on the functional equations for the 
transfer-matrices ${\Bbb T}$ to calculate the finite-size energy of 
the first excited state, the state which is interpreted as the 
one-particle state (with the zero-momentum particle) at $R\to \infty$ 
and approaches the CFT state associated with the 
identity operator of the CFT as $R\to 0$.

The paper is organized as follows. In Sect.2 we summarize the properties
of the operators ${\bf T}_j (\lambda)$ of the Conformal Theory
defined in \BLZ. The ``transfer-matrices'' ${\Bbb T}_j (\mu|\lambda)$ of 
the perturbed theory are introduced in Sect.3 where also our basic
conjectures about the analytic properties of these operators and their
relation to the local Integrals of Motion (IM) are formulated. In Sect.4
we study the simplest eigenvalues of the operator ${\bf
T}_{1/2}(\lambda)$ in the CFT ${\cal M}_{2/5}$. Similar analysis of the 
eigenvalues of ${\Bbb T}_{1/2}(\mu|\lambda)$ in SLYM is carried out in
Sect.5 where we derive the excited state energy $E_1(R)$. In Sect.6
the relation to the sine-Gordon theory is discussed and further 
applications of our approach are outlined.

\newsec{ The operators ${\bf T}_j (\lambda)$ in CFT}

In this section we briefly summarize the properties of the operators
$ {\bf T}_j (\lambda)$ in CFT as they are defined in \BLZ.

The space of states in a CFT decomposes as
\eqn\hcft{{\cal H}_{CFT} = \oplus_a \big( {\cal V}_{\Delta_a} \otimes 
{\bar {\cal V}}_{\Delta_a}\big)\ ,}
(for simplicity we consider here the ``diagonal" CFT) where
${\cal V}_{\Delta_a}$ is the irreducible highest weight representation 
of the ``left" Virasoro algebra $Vir$
\eqn\vir{
[L_n , L_m ] = (n - m)\, L_{n+m} + {c\over 12}\,
(n^3 - n)\,  \delta_{n+m,0}\ ,}
with the highest weight $\Delta_a$. Correspondingly, 
${\bar {\cal V}}_{\Delta_a}$ 
stands for the representation space of the ``right'' Virasoro algebra 
$\overline{Vir}$ identical to $Vir$.  The number $c$ 	in \vir\ is known 
as the ``central charge". The ``left" Virasoro generators $L_n$ are 
expressed in terms of the ``left" chiral component of the energy-momentum
tensor $T(z)$ ($z=x+iy$) as
\eqn\loim{ L_n = {c\over 24}\,  \delta_{n,0} + {R\over {2\pi}}\ 
\int_{0}^{R} {{dx}\over {2\pi}}\ 
T(x+iy)\ \ e^{i{{2\pi}\over R}n x}\ .}
Here we assume that the CFT is defined in the geometry of an Euclidean
cylinder so that $y$ is interpreted as the Euclidean time and the 
spatial coordinate $x$ is compactified on a circle of the circumference
$R$. The field $T(z)$ therefore satisfies the periodicity
condition $T(z+R) = T(z)$. The ``right" Virasoro generators
${\bar L}_n$ are defined in terms of the ``right" chiral component
${\bar T}({\bar z}),
\  {\bar z} = x - iy$ by similar integrals.

Any CFT possesses infinitely many local Integrals of 
Motion (IM) ${\bf I}_{2k-1}$ \jap, \Eguchi\ which can be written as
\eqn\loim{{\bf I}_{2k-1} = \int_{0}^{ R} {{dx}\over {2\pi}}\ 
T_{2k}(x+iy)\ ,}
where $T_{2k}(z)$ are certain local fields, polynomials in $T(z)$ and 
its derivatives. For example
\eqn\locdens{\eqalign{T_2 (z) = T&(z), \  T_4 (z) = :T^2 (z):, \  
T_6 (z) = :T^3 (z): + {{c+2}\over 12} :\big(T'(z)\big)^2 :,\ \ldots, \cr 
&T_{2k}(z) = : T^k (z):\  + \ {\rm terms\  with\  the\
derivatives}\ , }}
Here $:\ \ :$ denote appropriately regularized operator products, see
\ \BLZ\ for details.
There exists infinitely many densities \locdens\  (one
for each integer $k$ \Frenk) such that all
the integrals \loim\  commute
\eqn\locomm{[\, {\bf I}_{2k-1}, {\bf I}_{2l-1}] = 0\ .}

The ``transfer-matrices" ${\bf T}_j$ 
are most conveniently defined in
terms of Feigin-Fuchs free field
representation of the Virasoro algebra\ \Fei,\ \DotsFat.
Introduce the ``left" chiral Bose field\foot{
This field differs in normalization from the field $\varphi$ which
is used in \ \BLZ, $\varphi(z) = i\, \phi(z)$.}
\eqn\ffield{\phi(z) =   Q +    {{2\pi z}\over R}\ P  
-i\,  \sum_{n\neq 0}\
{{a_{-n}}\over n}\, e^{i{{2\pi} \over R}{nz}}\ \ ,} 
where the operators $Q, P, a_n$ satisfy the
canonical commutation relations
\eqn\cancomm{[Q,P] = {i\over 2} \, \beta^2 ,
\qquad [a_n , a_m ] = {n\over 2}\, \beta^2\, \delta_{n+m, 0} }
and $\beta$ is a parameter which will be defined below. Also, let 
${\cal F}_p$ be the Fock space, i.e the space generated by the action
of $a_n$ with $n < 0$ on the ``vacuum" state $\mid p \rangle$
which satisfies 
the equations
\eqn\vacdef{P\, \mid p \rangle = p\, \mid p \rangle\ ,
   \qquad a_n \mid p \rangle = 0 \quad { \rm for} \quad n>0\ .} 
Note that \ffield\  implies that the field $\phi(z)$ is quasi-periodic,
$\phi(z+R) = \phi(z) + 2\pi \,  P$. As is well known the 
energy-momentum tensor 
\eqn\ffuchs{T(z)=\beta^{-2}\ :(\partial_z \phi)^2 : +i \ (1-\beta^{-2})\ 
\partial_{z}^2 \phi - {1\over 24} }
generates the Virasoro algebra \vir\   with the central charge 
\eqn\cbeta{c = 13 - 6\, (\beta^2 + \beta^{-2})\ ,}
which therefore is represented in the Fock space ${\cal F}_p$. We will
always assume that $c < 1$ and hence that $\beta$ is real. For 
generic values of $\beta$ and $p$ the space ${\cal F}_p$ is 
isomorphic to the irreducible highest weight module ${\cal V}_{\Delta}$
with
\eqn\pdelta{
\Delta = \Delta(p) \equiv \Big( {p\over \beta}\Big)^2 + {c-1\over 24}\ ,}
while for particular values of these parameters, when the ``null-vectors"
appear (so called ``degenerate representations"), the irreducible 
representation ${\cal V}_{\Delta}$ can be 
obtained from ${\cal F}_p$ by factoring out the corresponding 
invariant subspaces. In fact, in what follows we always will imply that 
all such an invariant subspaces in ${\cal F}_p$
are factored out and
${\cal V}_p$ will  stand for the ``Virasoro irreducible
Fock space'', ${\cal V}_p = {\cal F}_{p}\,/\,{\hbox{(invariant
subspaces)}}$. With this convention the space ${\cal V}_p$ is  
isomorphic to ${\cal V}_{\Delta(p)}$. 

Let $j$ be a non-negative integer or half-integer number. Consider the 
following $(2j+1)\times (2j+1)$ matrix \BLZ
\eqn\Ljdef{{\bf L}_j (\lambda) = e^{i\pi P\, H_j} \ {\cal P} \exp \bigg(
\lambda \ \int_{0}^{R}  dz\,  \big( :e^{-2i\phi(z)}: q^{{H_j}\over 2} E_j +
:e^{2i\phi(z)}: q^{-{{H_j}\over 2}} F_j \big) \bigg)\ ,}
where $E_j , \ F_j$ and $H_j$ stand for the $2j+1$ dimensional
representation\ $\pi_j$\ 
of the quantum algebra $U_q \big(sl(2)\big)$ with
\eqn\qbeta{q = e^{i\pi\beta^2}\ ,}
and the symbol ${\cal P}$ in \Ljdef\   denotes the ``path ordering" of the 
both operator and matrix products. The matrix elements $\big[{\bf L}_j
(\lambda)\big]_{A}^{B} , \ A,B = -j, -j+1, ... , j$
are operators which
act in the space
\eqn\Fhat{
{\hat {\cal F}}_p =
\oplus_{n=-\infty}^{\infty} {\cal F}_{p+n\beta^2}\ .}
Note that the exponential fields $:e^{\pm 2i\phi(z)}:$ in \Ljdef\   have 
anomalous dimensions and therefore the parameter $\lambda$ carries the 
dimension
\eqn\dimlambda{
\lambda \sim \big[\, length\,\big]^{\beta^2 - 1}\ .}
As discussed in \ \BLZZ, the expression
\Ljdef\   applies directly only for
$\beta^2 < 1/2$ (i.e. for $c < -2$); for $\beta^2 > 1/2$ the integral in
\Ljdef\   can be defined through appropriate analytic continuation in 
$\beta^2$.

The operator matrices ${\bf L}_j (\lambda)$ satisfy the Yang-Baxter 
relation
\eqn\YBaxter{\eqalign{
&\sum_{C=-j}^{j}\ \sum_{C'=-j'}^{j'}\ 
\big[R_{j\,j'}(\lambda/\lambda')\big]_{A\,A'}^{C\,C'}\ \big[{\bf L}_j
(\lambda)\big]_{C}^{B}\ \big[{\bf L}_{j'} (\lambda')\big]_{C'}^{B'} =\cr
&\sum_{C=-j}^{j}\ \sum_{C'=-j'}^{j'}\ \big[{\bf L}_{j'}
(\lambda')\big]_{A'}^{C'} \ \big[{\bf L}_j (\lambda)\big]_{A}^{C} 
\ \big[R_{j\,j'}(\lambda/\lambda')\big]_{C\,C'}^{B\,B'}\ ,}}
where $\big[R_{j\,j'}(\lambda)\big]_{A\,A'}^{B\,B'}$
are the elements of the
$R$-matrix acting in the product of $2j+1$ and $2j' +1$ dimensional
representations of $U_q \big(sl(2)\big)$.
As a simple consequence of \YBaxter\ 
the ``transfer-matrices'' 
\eqn\Tjdef{{\bf T}_j (\lambda) =  
tr_{\pi_j} \big[\, e^{i\pi P H_j}\ {\bf L}_j (\lambda)\, \big] \equiv 
\sum_{A=-j}^{j}\big[\, e^{i\pi P H_j}\ {\bf L}_j (\lambda)\, \big]_{A}^{A}}
form a commuting family of operators, i.e. they satisfy \ \Tcomm. Note
that although the off-diagonal elements of the matrices \Ljdef\  act
between different components of \Fhat\  the operators \Tjdef\ 
invariantly act in the spaces ${\cal V}_p$,
\eqn\Tjact{
{\bf T}_j (\lambda) : \qquad {\cal V}_p  \to  {\cal V}_p\ .} 

It is possible to show using \Ljdef,\ \YBaxter\ 
that the operators ${\bf T}_j
(\lambda)$ satisfy the fusion relations \Tfusion
\ \BLZ,\ \BLZZZ\  which allow one 
to express any operator ${\bf T}_j (\lambda)$ with $j > 1/2$ through 
the ``fundamental" one  ${\bf T}(\lambda)\equiv {\bf T}_{1/2}(\lambda)$. 

The operator ${\bf T}(\lambda)$ is an entire function of $\lambda^2$ (the 
same is true for all the operators ${\bf T}_j (\lambda)$). Its power
series expansion in the variable\ $\lambda^2$\ 
follows directly from the definitions \Ljdef, \Tjdef\ 
\eqn\tseries{{\bf T}(\lambda) =
2\ \cos(2\pi P) + \sum_{k=1}^{\infty} \lambda^{2 k}\, 
{\bf G}_k\ ,}
where ${\bf G}_k$ are basic ``nonlocal IM"
\eqn\nloim{\eqalign{
{\bf G}_k = q^k\,
\    & \int_{{\cal D}_R(x_1,...,x_{2k})}\Big(\ 
e^{2i\pi P}\ :e^{-2i\phi(x_1)}::e^{2i\phi(x_2)}:
:e^{-2i\phi(x_3)}:...:e^{2i\phi(x_{2 k})}: \cr &\ \ \ +
e^{-2i\pi P}\ 
:e^{2i\phi(x_1)}::e^{-2i\phi(x_2)}:
:e^{2i\phi(x_3)}:...:e^{-2i\phi(x_{2 k})}:\ \Big)\ .}}
Here the integration is performed over the domain $D_{R}(x_1,...,x_{2k}) 
= \{R > x_1 > x_2 > ... > x_{2k}>0 \}$. In the opposite limit $\lambda^2
\to \infty$ the operator ${\bf T}(\lambda)$ admits asymptotic expansion
\BLZ, \BLZZ\
\eqn\tass{\log {\bf T}(\lambda)\ 
\simeq \ ({\kappa\,R}/{2\pi})\  \lambda^{(1+\xi)}
\ {\bf I} -\sum_{n=1}^{\infty}\ C_n\  \lambda^{(1-2n)(1+\xi)}
\ {\bf I}_{2n-1}\ .}
in powers of $\lambda^{-(1+\xi)}$ 
with the coefficients proportional
to the local IM \loim. The numerical
factors $\kappa$ and $C_n$ are given by
\eqn\coeffcn{\eqalign{&
\kappa= {{2\sqrt\pi\  \Gamma({1\over 2} - {\xi\over 2})}\over 
{\Gamma(1-{\xi\over 2})}}\
\Big(\Gamma\big(1-\beta^2\big)\Big)^
{1+\xi}\ ,\cr  
C_n = {{\sqrt\pi (1+\xi)}\over {  n!}}\  & \big(\, \beta^{2}\,
\big)^n\  
{{\Gamma\big((n-{1\over 2})(1+\xi)\big)}\over 
{\Gamma\big(1+(n-{1\over 2})\xi\big)}}\
\Big(\Gamma(1-
\beta^2)\Big)^{-(2n-1)(1+\xi)}\ .}}
Here and below
\eqn\xibeta{
\xi = {\beta^2 \over {1-\beta^2}}\ .}

The above expansions suggest that the eigenvalues  of the
operator ${\bf T}(\lambda)$ unite all the information about the 
eigenvalues of both
nonlocal IM \nloim\  and local IM \loim\  in a
very interesting manner. In what follows we consider almost 
exclusively the vacuum eigenvalues
\eqn\Tvac{
{\bf T}(\lambda)\mid p \rangle  =
 T(t,p) \mid p \rangle \ ; \qquad
t\equiv \lambda^2 \, \Big(\, {R\over 2\pi}\, \Big)^{2-2\beta^2}\ .}
According to \tseries\  and \tass\  this function expands both in 
convergent series around $t =0$ and in asymptotic series in the 
vicinity of $t = \infty$,
\eqna\eigenexp
$$\eqalignno{
T(t,p)& = 2\cos (2\pi p) + \sum_{k=1}^{\infty} t^k\ 
G^{vac}_k (p)&\eigenexp a\cr
& \simeq \exp\Big( \kappa \  t^{{1+\xi}\over 2}
\ -\sum_{n=1}^{\infty}\ C_n\ t^{(1/2-n)(1+\xi)}
\ {I}^{vac}_{2n-1}\big(\Delta(p)\big) \Big)\ ,&\eigenexp b\cr}$$
where $G^{vac}_k (p) $
and $I^{vac}_{2n-1}(\Delta)$
are, respectively, the vacuum eigenvalues of the nonlocal IM \nloim\  and the
local IM \loim\ with $R=2\pi$. Note, that the expression in
\eigenexp{b}\ 
is, in fact, an asymptotic series with zero radius of convergence. 
The eigenvalues $G^{vac}_k (p)$ are given by the integrals
\eqn\intgk{\eqalign{
G^{vac}_k& (p) =
\int_{0}^{2\pi} du_1 \int_{0}^{u_1} dv_1\int_{0}^{v_1} du_2
\int_{0}^{u_2} dv_2...\int_{0}^{v_{n-1}} du_{n}\int_{0}^{u_{n}} dv_{n}\ \cr 
&\prod_{j>i}^{n}\Biggr[{\Bigl(2\ \sin\big({{u_i - u_j}\over 2}\big)
\Bigl)}^{2\beta^2}\ 
{\Bigl(2\ \sin\big({{v_i - v_j}\over 2}\big)\Bigl)}^{2\beta^2}\Biggr]
\prod_{j\geq i}^{n}{\Bigl(2\ \sin\big({{u_i - v_j}\over 2}\big)
\Bigl)}^{-2\beta^2}\times\cr &
\prod_{j>i}^{n}{\Bigl(2\ \sin\big(
{{v_i - u_j}\over 2}\big)\Bigl)}^{-2\beta^2}\, 
 2\, \cos\Bigl(2 p\
\big(\pi+\sum_{i=1}^{n} (v_i - u_i ) \big)\Bigl)\ ,}}
while
the eigenvalues $I^{vac}_{2n-1}(\Delta)$
can be calculated (in principle) from the
explicit formulas for the associated local densities $T_{2n}(z)$ 
in \loim. Unfortunately for general $k$ neither the integrals  \intgk\ 
can be calculated
explicitly nor the densities $T_{2k}(z)$ are known in a 
closed form. Some partial results about
the vacuum eigenvalues $G^{vac}_k (p)$
and $I^{vac}_{2n-1} (\Delta)$
are collected in Appendices  A and B. 

For $\beta^2 < 1/2$ the entire function $T(t,p)$ is completely
determined by the positions $t_k$ of its zeroes in the complex 
$t$-plane. The locations of these zeroes are known exactly in two
particular cases. First, in the ``classical limit'' $\beta^2 \to 0$,
(with $p$\ and $\lambda$ kept fixed)   the
explicit formula for
the vacuum eigenvalue $T(t,p)$ can be easily  obtained
directly from the definition \Tjdef\foot{
To derive (2.28) one observes that the oscillator modes $a_n$ can be 
ignored in the vertex operators $:e^{\pm 2 i \phi}:$ as they do not 
contribute to the vacuum eigenvalues in this limit. Also, in this limit
the zero mode operators $P$ and $Q$ commute and one obtains
$T_j(t,p)=tr_j\Big( e^{2\pi\, (\sqrt{t}E_j+\sqrt{t}F_j+ipH_j)}\Big).$}
\eqn\Tclass{
T(t,p) = 2\ \cos\big(
2\pi \sqrt{p^2 - t}\,\big); \qquad \beta^2 = 0\ .}
In this case
\eqn\tkclass{
t_k = p^2 - {(2 k+1)^2\over 16} \ , \quad k=0, 1, 2, ... .}
Another case where $T(p,t)$ is known explicitly is the ``free-fermion
point'' $\beta^2 = 1/2$ \ \BLZZ,
\eqn\Tff{T(t,p) = {{2\pi\  e^{2{\cal C}t}}\over 
{\Gamma(1/2+2p+\pi t)\,
\Gamma(1/2-2p+\pi t)}}, \qquad \beta^2 = 1/2\ ,}
where ${\cal C}$ is some  constant.
This function has two sequences of zeroes
\eqn\tkff{t_{k}^{(\pm)} = {1\over \pi}\ (\pm 2p - 1/2 - k) \ , \quad 
k=0, 1, 2, ...\  .}
For generic $0<\beta^2<{1\over 2} $ the zeroes $t_k$ do not have any 
simple form. Nonetheless a brief inspection of \eigenexp a\ and
\tkclass\ ,  \tkff \  makes it natural to adopt the 
following

\vskip 0.1in

{\bf Conjecture 1.}
{\it For real $p$ and 
$0 \leq \beta^2 \leq 1/2$ all zeroes $t_k$ of $T(t,p)$ are
real. There are infinitely many negative zeroes which accumulate towards
$t = -\infty$; for given value $p$ there are also exactly \/${\rm Int}
(2|p| +1/2)$ 
positive zeroes (\/${\rm Int}(x)$ denotes integer part of $x$). In particular, 
if\  $|p| < 1/4$ all zeroes $t_k$ are negative.}

This assumption will be used in Sect.4 in deriving closed nonlinear
integral equation for the vacuum eigenvalues in particular case $\beta^2
= 2/5$.

\newsec{ The Operators ${\Bbb T}_j (\mu|\lambda)$ in the $\Phi_{1,3}$
Perturbed Theory}

Our previous discussion was concerned with the Conformal Field 
Theory. It is well known that one can obtain more general non-
conformal Integrable Quantum Field Theory by perturbing certain
CFT with appropriately chosen relevant local operator. The most known 
example is a ``minimal CFT'' perturbed with the degenerate field
$\Phi_{1,3}$, i.e. the primary field with the conformal dimensions
\eqn\del{\Delta_{1,3} = {{\bar \Delta}}_{1,3} = -1+2\beta^2 =
1-{2\over{1+\xi}}\ ,}
which satisfy the third-level null vector equations\  \BPZ
\eqn\nullvector{\eqalign{&
\big(2\, (1+2\beta^2)\,L_{-3} - 4\, L_{-1}L_{-2}
 +\beta^{-2}\,  L_{-1}^3 \big)\ \Phi_{1,3}(z, {\bar z}) = 0\ ,\cr &
\big(2\, (1+2\beta^2)\,{\bar L}_{-3} - 4\, 
{\bar L}_{-1}{\bar L}_{-2} + \beta^{-2}\,{\bar L}_{-1}^3 \big)\ 
\Phi_{1,3}(z, {\bar z}) = 0\ .}}
We will assume the canonical normalization of this field
\eqn\phinorm{\langle \Phi_{1,3}(z,{\bar z})\, \Phi_{1,3}(z',{\bar z}')
\rangle \to \big[(z-z')({\bar z} - {\bar z}')\big]^
{-2\Delta_{1,3}}\quad {\rm as} \quad (z,{\bar z})\to (z',{\bar z}')\ .} 
The perturbed theory is defined by the formal action
\eqn\action{{\cal A}_{PCFT} = {\cal A}_{CFT} + {\hat \mu}^2 \,
\int \,d^2 x 
\ \Phi_{1,3}(z,{\bar z}) \ ,}
where ${\cal A}_{CFT}$ denotes the ``action''
of CFT and ${\hat \mu}$ is the
coupling parameter with the dimension 
\eqn\dimmmu{{\hat \mu} \sim \big[\, length\, \big]^{\Delta_{(1,3)}-1} =
\big[\, length\, \big]^{2\beta^2 - 2}\ .}
As in Sect.2 we assume that the QFT is defined in the geometry of a 
cylinder with the spatial coordinate $x = \Re e\, z$ compactified on 
a circle with the circumference $R$. Correspondingly, the field
$\Phi_{1,3}$ in \action\  is taken to be periodic function of $x$, i.e.  
$\Phi_{1,3}(z+R,{\bar z}+R) = \Phi_{1,3}(z,{\bar z})$.
As was shown in\  \ABZ\
the QFT defined by \action\  is integrable in the 
sense that it possesses infinitely many local IM which appear as certain
deformations of the local IM \loim\  of the original CFT. Although for 
${\hat \mu} \neq 0$ the fields
$T_{2k}$ \locdens\  are no longer holomorphic
fields  (i.e. they depend both on $z$ and ${\bar z}$) they satisfy the
continuity condition
\eqn\continuity{
\partial_{\bar z} T_{2k}(z,{\bar z}) =
{\hat \mu}^2 \,\partial_z \Theta_{2k-2}
(z, {\bar z})\ ,}
where $\Theta_{2k-2}(z,{\bar z})$ are certain local fields. The
right-moving chiral sector of the original CFT contains also the
antiholomorphic 
fields ${\bar T}_{2k}$; in the perturbed theory \action\ 
they satisfy similar continuity equations
\eqn\continuityo{\partial_ {z}
{\bar T}_{2k}(z, {\bar z}) ={\hat  \mu}^2
 \,\partial_{\bar z} 
{\bar \Theta}_{2k-2} (z, {\bar z})\ .}
It follows that the integrals
\eqn\pertloim{\eqalign{&
{\Bbb I}_{2k-1} = 
\int_{0}^{ R}\, {d x\over 2\pi}\,  \big(T_{2k}(x+iy,x-iy) +
{\hat  \mu}^2 
\,\Theta_{2k-2} (x+iy,x-iy)\big)\ ,\cr
&{\bar{\Bbb I}}_{2k-1} = \int_{0}^{ R}\, {d x\over 2\pi}\, 
\big({\bar T}_{2k}(x+iy,x-iy)
+{\hat  \mu}^2 \,{\bar \Theta}_{2 k-2}(x+iy,x-iy)\big)}}
in fact do not depend on the ``Euclidean time'' $y$, i.e. they are 
Integrals of Motion (IM). It is possible to show \ \ABZ\ 
that the operators
\pertloim\  commute among themselves
\eqn\loimcomm{[{\Bbb I}_{2k-1},{\Bbb I}_{2l-1}] = [{\Bbb I}_{2k-1},
{\bar {\Bbb I}}_{2l-1}] = [{\bar {\Bbb I}}_{2k-1},
{\bar {\Bbb I}}_{2l-1}] = 0\ .}
Note that the field 
${\hat \mu}^2 \,\Theta_0 ={\hat  \mu}^2 \,{\bar \Theta}_0 =
{\hat  \mu}^2
\, (1-\Delta_{1,3})\, \Phi_{1,3}$ coincides
with the trace $T_{\alpha}^{\alpha}$ of the energy-momentum tensor
$T_{\alpha \beta}$ of the QFT \action, and the IM ${\Bbb I}_1$ and
${\bar {\Bbb I}}_1$ are related to the Hamiltonian ${\Bbb H}$ and the
spatial momentum ${\Bbb P}$ of this QFT
\eqn\hamilton{{\Bbb H}={\Bbb I}_1 + {\bar {\Bbb I}}_1 , \qquad 
{\Bbb P}={\Bbb I}_1 - {\bar {\Bbb I}}_1\ .}

As in Sect.2 it is convenient to use the 
Feigin-Fuchs free-field realization for the original CFT.
The space
of states \hcft\  can be realized as 
\eqn\hcfto{{\cal H}_{CFT} = {\oplus_{p}}\, {\cal H}_p ; \qquad
{\cal H}_p \equiv {\cal V}_{p} \otimes {\bar {\cal V}}_{-p}\ ,} 
where ${\cal V}_{p}$ are the ``irreducible Fock spaces'' defined
in Sect.2 and ${\bar {\cal V}}_{-p}$ are  similar spaces 
generated by the ``right-moving'' chiral Bose field
\eqn\ffieldo{
{\bar \phi}({\bar z}) =
{\bar Q} -\, {{2\pi \bar z}\over R}{\bar P} -i\, 
\sum_{n\neq 0}\, {{{\bar a}_{-n}}\over n}\,
e^{-i{{2\pi}\over R}{n \bar z}}\ .}
Each of the spaces ${\cal H}_p$ 
forms an irreducible representation
of $Vir\oplus {\overline{Vir}}$
with the highest weight $\big(\Delta(p),
\Delta(p)\big)$ and the highest weight vector 
\eqn\erw{\mid \Delta(p)\rangle
= \mid p\rangle\otimes{\overline{\mid -p\rangle }}\ ,}
where $\mid p\rangle$ and $\overline{\mid- p\rangle}$
are the vacuum vectors in ${\cal F}_p$ and ${\bar{\cal F}}_{-p}$, 
respectively. The sum in \hcfto\  is taken over certain set 
of admitted values of the parameter $p$ which is specific
for each CFT and  determines its Virasoro representation content. In
different theories this set can be discrete or continuous.

One remark is in order here. The above perturbed QFT \action\  can be
constructed only if the original CFT meets certain criteria. For 
instance, as in CFT the space of local fields is isomorphic to the
space of states \ \hcfto , the sum in \hcfto\  must contain the term 
with $2 p=3\beta^2-1$ corresponding to the field $\Phi_{1,3}$ (and its
conformal descendants). Moreover, according to the fusion rules which
follow from the null-vector equations \nullvector\  the operator 
$\Phi_{1,3}$ acts as
\eqn\phiact{
\Phi_{1,3} \ : \quad  {\cal H}_{p}  \to  {\cal H}_
{p-\beta^2} \, 
\oplus \, {\cal H}_p \, \oplus \, {\cal H}_{p+\beta^2} \ ,} 
and therefore to admit the action of this operator the space \hcfto\
must contain the whole sum
\eqn\hhat{{\hat{\cal H}}_p = \oplus_n\,  {\cal H}_{p+n\beta^2} }
together with each ${\cal H}_p$
included therein. For particular values of
$p$ (corresponding to so called ``completely degenerate''
representations of $Vir$) this sum truncates and contains in fact  
only finitely many terms (see \ \BPZ).
In any case we will assume that
the space \hcfto\  has the structure
\eqn\hcftoo{{\cal H}_{CFT} = \oplus_{a}\,  {\hat {\cal H}}_{p_a}\ .}
Obvious example of CFT with this structure is ``minimal CFT'' ${\cal
M}_{r/r'}$ with $\beta^2={r\over r'}$
\ \BPZ\ in which case $p$ in \hcfto\  run over finitely many 
values $2\, p_{l,k} =
l - k\, \beta^2\ ( l=1,2,...,r-1;\  k=1,2,...,r'-1)$. 
Another example is $c=1$ CFT
of uncompactified free Bose field where
$p$ in \hcfto\  takes continuous values.

The perturbation term in \action\ 
is relevant (i.e. $\Delta_{1,3} <
1$) for  $0<\beta^2 < 1$ and so the perturbation theory in
$\mu^2$ is superrenormalizable (moreover, for $\beta^2 < 1/2$ this
perturbation theory does not have any short-distance divergences at
all). Therefore in the finite-size system with $R<\infty$ where possible
infrared divergences are also well contained there are all reasons to
assume that the space of states ${\cal H}_{PCFT}$ of the perturbed
theory \action\  coincides with the space of states \ \hcft\  of the
original CFT
\eqn\HsimH{{\cal H}_{PCFT} \simeq {\cal H}_{CFT}\ .}
Of course the highest weight states $\mid \Delta(p) \rangle$ \erw\ 
in ${\cal H}_{CFT}$ 
are no longer stationary states of the perturbed Hamiltonian
\hamilton; finding the eigenstates and eigenvalues of ${\Bbb H}$ in
\HsimH\  is exactly the problem we want to address here. 

In Sect.2 we have defined the operator valued matrices 
${\bf L}_j (\lambda)$ for the ``left'' chiral sector of the CFT; these
matrices satisfy the Yang-Baxter relation \YBaxter. Similar way one can
define the ``right'' chiral counterparts of these matrices,
\eqn\Ljbar{{\bar {\bf L}}_j (\lambda) = 
\bigg[{\cal P}\exp \bigg(\lambda \int_{0}^{R} \big(:e^{-2i{\bar \phi}
(\bar z)}:
\, q^{{H_j}\over 2} E_j +
:e^{2i{\bar \phi}(\bar z)}:\,q^{-{{H_j}\over 2}} F_j \big)\,d{\bar
z}\bigg)\bigg]\ e^{-i\pi{\bar P}\, H_j}\  ,}
where ${\bar\phi}({\bar z})$ is the right chiral Bose field \ffieldo,
$E_j,\ F_j$ and $H_j$ denote 
the same $(2j+1)\times (2j+1)$ matrices as
in \Ljdef, and like in \Ljdef\  ${\cal P}$ 
indicates the path ordering of both 
operators and matrices along the integration path. The elements of 
the matrices \Ljbar\
act in the space ${\hat{\bar{\cal F}}}_p$ defined 
similarly to \Fhat\  
in terms of the ``right'' Fock spaces ${\bar {\cal
F}}_p$. It is possible to check that the operator matrices \Ljbar\  
also satisfy the Yang-Baxter relation in the form
\eqn\YBaxterbar{\eqalign{&\sum_{C=-j}^{j}\ \sum_{C'=-j'}^{j'}\ 
\big[{\bar {\bf L}}_j (\lambda)\big]_{A}^{C}\  
\big[{\bar {\bf L}}_{j'} (\lambda')\big]_{A'}^{C'}\ 
\big[R_{j\,j'}(\lambda/\lambda')\big]_{C\,C'}^{B\,B'} =\cr
&\sum_{C=-j}^{j}\ \sum_{C'=-j'}^{j'}\ 
\big[R_{j\,j'}(\lambda/\lambda')\big]_{A\,A'}^{C\,C'}\ 
\big[{\bar {\bf L}}_{j'} (\lambda')\big]_{C'}^{B'}\ 
\big[{\bar {\bf L}}_j (\lambda)\big]_{C}^{B}\ ,}} 
where $R_{j\,j'}(\lambda)$ are exactly the same $R$-matrices as in
\YBaxter. Note that the $R$-matrices enter \YBaxterbar\
different way as
compared to \YBaxter, the difference being traced down to the 
difference in the commutation relations
\eqn\phicomm{[\phi (x), \phi(x')] = - [{\bar \phi} (x), {\bar \phi}(x')] 
= {\pi \beta^2\over 2 \, i}\ 
\varepsilon \Big({{x-x'}\over R}\Big)\ ,}
where $\varepsilon(x)$ is the quasi-periodic continuation of the 
${\rm sign}$ function,
\eqn\vareps{\epsilon(x) = 2 n+1,   \quad n<x<n+1 \ ;\  n \in
{\Bbb Z}\ .}

Now we can introduce the following ``full transfer-matrices''
\eqn\fullT{{\Bbb T}_j (\mu |\lambda) =
{ tr_{\pi_j}}\Big[\, {\bf L}_j (\lambda) 
\, {\bar {\bf L}}_j (\mu/\lambda)\, \Big] = \sum_{A,B=-j}^{j}
\ \big[{\bf L}_j (\lambda)\big]_{A}^{B}\ \big[{\bar {\bf L}}_j 
(\mu/\lambda) \big]_{B}^{A}\ ,}
where $\mu$ is a parameter.
It is easy to check that \fullT\  defines the operator which acts in
the space \hhat, i.e.
\eqn\fullTact{{\Bbb T}_j (\mu |\lambda)\ : \qquad {\hat {\cal H}}_p  \to  
{\hat {\cal H}}_p\  .}
It follows from \YBaxter,  \YBaxterbar\  and the fact that the
$R$-matrices satisfy the ``unitarity conditions''
\eqn\unitarity{\sum_{C=-j}^{j}\ \sum_{C'=-j'}^{j'}\big[
R_{j\,j'}(\lambda)\big]_{A\,A'}^{C\,C'}\
\big[R_{j\,j'}(\lambda^{-1})\big]_{C\,C'}^{B\,B'} = \delta_{A}^{B}\ 
\delta_{A'}^{B'}\ ,}
that these operators form the commuting family, i.e.
\eqn\fullcomm{
[{\Bbb T}_j (\mu|\lambda), {\Bbb T}_{j'} (\mu| \lambda')] = 0\ } 
for all values of $j, j'$. The following simple properties of the
operators  \fullT\  are immediately established
\vskip 0.1in

1. For $\mu = 0$ the operator ${\Bbb T}_j$ reduces to the operator
${\bf T}_j$ defined in Sect.2, 
\eqn\zeromu{{\Bbb T}_j (0|\lambda) =
 {\bf T}_j (\lambda) \otimes {\bar {\bf I}}\ ,}
i.e. it acts as ${\bf T}_j (\lambda)$ in the ``left''  spaces
${\cal V}_p$ and as the identity operator ${\bar {\bf I}}$ in the 
``right''  spaces ${\bar {\cal V}}_p$ in \hcfto\foot
{Obviously, the zero
mode operator ${\bar P}$, when acting on the space
\hcfto, can be identified \hfill\break with $-P$.}.
Similarly,
\eqn\zeromubar{
{\Bbb T}_j (\mu|\mu/{\lambda})\big|_{\mu = 0} = {\bf I}
\otimes {\bar {\bf T}}_j ({\lambda})\ ,}
where ${\bf I}$ is the identity in the ``left'' spaces ${\cal V}_p$
and ${\bar {\bf T}}_j (\lambda) =  tr_{\pi_j}\big[\,
{\bar {\bf L}}_j (\lambda)\, e^{-i\pi {\bar P}H_j}\,\big]$
act in the spaces ${\bar{\cal V}}_p$ as 
the ``right'' counterparts of the operators \Tjdef.

\vskip 0.1in

2. Like ${\bf T}_j (\lambda)$ the operators ${\Bbb T}_j (\mu|\lambda)$ 
are the single-valued functions of $\lambda^2$.

\vskip 0.1in

3. As follows directly from the definition \fullT\  the operators 
${\Bbb T}_j (\mu |\lambda)$ inherit their large $\lambda$ asymptotics from
those of the operators ${\bf T}_j (\lambda)$,
\eqn\fullass{\log {\Bbb T}_j (\mu|\lambda) \sim {\kappa}_j\, R/(2\pi)
\ \lambda^{1+\xi}, 
\qquad \lambda  \to  \infty\ ,}
where
\eqn\kappaj{
{\kappa}_j = {{\sin(\pi j \xi)}\over \sin(\pi\xi/2)}\ \kappa\ } 
and $\kappa$ is given by \coeffcn. Similarly, the asymptotics of these 
operators at $\lambda \to 0$ are controlled by the second factor in 
\fullT\   and so
\eqn\fullasss{\log {\Bbb T}_j (\mu |\lambda) \sim {\kappa}_j\, R/(2\pi)\  
(\mu/\lambda)^{1+\xi}, \qquad \lambda \to  0\ .} 
In particular, it follows from \fullasss\  that ${\Bbb T}_j 
(\mu |\lambda)$ has an essential singularity at $\lambda^2 = 0$.

It is clear that for generic value of the parameter $\mu$ the operators
${\Bbb T}_j (\mu |\lambda)$ do not commute with the CFT local IM \loim.
Instead, it is natural to expect that the operators ${\Bbb T}_j (\mu
|\lambda)$ play the same role in perturbed theory \action\  as ${\bf T}_j
(\lambda)$ did in the CFT, namely, the asymptotic expansions of ${\Bbb
T}_j (\mu |\lambda)$ near the essential singularities at $\lambda^2 =0$
and $\lambda^2 = \infty$ generate the local IM \pertloim\  of the
perturbed theory \action. Note that the spaces \hhat\  of the perturbed 
theory admit the action of the perturbed local IM \pertloim. The
parameter $\mu$ in \fullT\  is expected to be related to the coupling 
parameter ${\hat \mu}$ in \ \action.

Concerning the operators ${\Bbb T}_j$ we can formulate now our basic
\vskip 0.2in

{\bf Conjecture 2}
{\it
\vskip 0.1in

a) The operators ${\Bbb T}_j (\mu |\lambda)$ are single-valued 
functions of 
$\lambda^2$, regular everywhere in the complex $\lambda^2$ plane except
at $\lambda^2 = 0$ and $\lambda^2 = \infty$ where they have essential
singularities.

\vskip 0.1in

b) The asymptotic expansions of the
operators ${\Bbb T}_j (\mu |\lambda)$
near $\lambda^2 =
\infty$ and $\lambda^2 = 0$ are controlled by the 
perturbed local IM \pertloim; in particular, for ${\Bbb T}(\mu
|\lambda) \equiv {\Bbb T}_{1/2} (\mu |\lambda)$ one has
\eqn\fullTexp{\eqalign{&\log {\Bbb T} (\mu |\lambda) =
\kappa\,R/(2\pi) \
\lambda^{1+\xi} - 
\sum_{n=1}^{\infty}\ C_n \ \lambda^{(1-2n)(1+\xi)}\ {\Bbb I}_{2n-1}\ ,\cr
&\log {\Bbb T} (\mu |\lambda) =
\kappa\,R/(2\pi)\ (\mu/\lambda)^{1+\xi} - 
\sum_{n=1}^{\infty}\ C_n \ \big(\mu/\lambda \big)^{(1-2n)(1+\xi)}\ {\bar 
{\Bbb I}}_{2n-1}\ ,}}
where $\kappa$ and $ C_n$ are exactly
the same constants as in \coeffcn\ and
the parameter $\mu$  is related to   ${\hat \mu}$ 
in \action\ as
\eqn\rel{{\hat \mu}^2={\Gamma^4\big(1-\beta^2)\over 
\pi\,(1-2\beta^2)(3\beta^2-1)}\  \biggl({\Gamma\big(3\beta^2\big)\, 
\Gamma\big(\beta^2\big)\over \Gamma\big(1-3\beta^2\big)\,
\Gamma\big(1-\beta^2\big)}\biggr)^{1\over 2}\ \mu^2 .}
In particular, the ``transfer-matrices'' ${\Bbb T}_j$ commute with
all local IM \pertloim
\eqn\tloimcomm{[{\Bbb T}_j (\mu|\lambda),{\Bbb I}_{2n-1}] = 
[{\Bbb T}_j (\mu|\lambda) , {\bar {\Bbb I}}_{2n-1}] = 0\ .}
}
\vskip 0.1in

The relation \rel\ appears very natural in view of nearly obvious
relation of the operators ${\Bbb T}(\mu|\lambda)$ to the sine-Gordon
model (which we nonetheless discuss briefly in Sect.6 below); we have
arrived at \rel\ by combining \coeffcn, \fullass-\fullasss\  with known
relations (6.3), (6.4) and (6.6).
  
It is possible to show that the operators \fullT\
satisfy exactly the same ``fusion relations" as the 
operators ${\bf T}_j$, namely
\eqn\fullfusion{
{\Bbb T}_j (\mu|q^{1\over 2}\lambda) {\Bbb T}_j (\mu|q^{-{1\over 2}}
\lambda) = {\Bbb I} + {\Bbb T}_{j+1/2}(\mu|\lambda){\Bbb T}_{j-1/2}
(\mu|\lambda)\ .}

Also, if $q$ is a root of unity the fusion
relations \fullfusion\  truncate
exactly the same way as it
happens in the unperturbed theory \ \BLZ\  
(see  Sect.4 below). This
leads to closed functional equations for
the eigenvalues of the operators
${\Bbb T}_j (\mu |\lambda)$
in the space ${\cal H}_{PCFT}$. Solving
these equations one can, in principle, determine those
eigenvalues and whence find the eigenvalues of all the local IM
\pertloim\ in the perturbed
theory \action, including the eigenvalues 
of the Hamiltonian \hamilton. In Sect.5
we will show how this
approach works in the SLYM \SLYM. 

\newsec{Fussion Truncation and Kl\"umper-Pearce Equations in CFT}

The properties of the operators ${\bf T}_j$ described Sect.2. hold 
for arbitrary positive $\beta^2$. As was discussed in
\ \BLZ\  considerable
simplifications occur 
if $\beta^2$ is a rational number (in this case 
$q$ defined in \qbeta\  is a root of unity and associated 
value of $c$ corresponds
to a ``minimal CFT" \ \BPZ ). In these cases the 
``fusion relation" 
\Tfusion\  truncates and reduces to a functional relation
involving finitely many
operators ${\bf T}_j$. This ``truncated" fusion
relation leads then
to the closed system of functional equations for the
eigenvalues $T_j (\lambda)$ of these operators. 

Let $q$ be a complex number of the form
\eqn\qN{q^{N}=\pm 1\, , \qquad {\rm and}\qquad  q^n\neq\pm1\  \ 
{\rm for\  any \ integer }\ 0<n<N\ ,}
where $N\ge2$ some integer.
%
%
For such values of $q$ the $N+1$ dimensional representation  
of $U_q \big(sl(2)\big)$ with the spin $j=N/2$ becomes reducible,
while all the finite dimensional representation with $j<N/2$ remain 
irreducible.
More precisely,
this $N+1$ dimensional representation reduces to
a direct sum
of the  $N-1$ dimensional representation
with $j = N/2-1$ and two
one-dimensional representations \foot{The latter are 
special one-dimensional  representations  $\rho_{\pm}$  of
 $U_q \big(sl(2)\big)$ with
$q^{2N}=1$ for which $\rho_{\pm}(E)=\rho_{\pm}
(F)=0$ and $\rho_{\pm}(H)=\pm N$.}.
These representations  bring
separate contributions to the trace  \Tjdef\  and therefore
\eqn\Tred{{\bf T}_{N/2} (\lambda) = 2 \cos (2\pi N\/ {\bf P}) + 
{\bf T}_{N/2 -1}(\lambda)\ ,}
where we used the
notation ${\bf P}$ for $P$ to emphasis its operator
nature. Due to \Tred\  
the fusion relations \Tfusion\  close within $N$
operators ${\bf T}_j (\lambda)$ with $j =0, 1/2,1,...,N/2 -1/2$.

This truncated fusion
relation leads to closed functional equations for
the eigenvalues of
the operators ${\bf T}_j (\lambda);\ j = 0,1/2,1,...,
N/2-1/2$. Let $T_j (\lambda)$
be the eigenvalues of these operators associated
with some common eigenstate in $ {\cal V}_p$.
The functional equations take most convenient form
if one makes use of the rapidity variable
\eqn\thetak{\theta = \log\big(\, \lambda^{1+\xi} R/(2\pi) \,\big)\ ,}
and introduces the functions \ \KluP
\eqn\Yjdefa{\eqalign{Y_j (\theta) =
T_{j-\half}(\lambda)&\,T_{j+\half }(\lambda),
\quad j=\textstyle{\half ,1,{3\over 2},...,{N\over2}-1}\, ;\cr
Y_0 (\theta)& 
\equiv 0, \qquad \overline{Y}(\theta) = T_{{N\over2}-1}(\lambda)\ .}}
Then the functional equations can be written as
\eqn\DNtba{\eqalign{&
Y_j \Big(\theta + {i\pi\xi\over2}\Big)\,Y_j \Big(\theta - {i\pi  \xi\over2}
\Big) =
\big(1+Y_{j-\half }(\theta)\big)\big(1+Y_{j+\half }(\theta)\big),
\  j = \textstyle{\half , 1,\ldots, {N\over2}-{3\over2}} ,\cr 
&Y_{{N\over2}-1}\Big(\theta + {i\pi\xi\over2}\Big)\,
Y_{{N\over2}-1}\Big(\theta - {i\pi\xi\over 2}\Big)=\cr
&\ \ \ \ \ \ \ \ \ \ \ \ \ \ \ \ \ \ \ \ \ \ \ \ \ \ \ \ \
=\big(1+Y_{{N\over2}-{3\over2}}(\theta)\big)\,
\big(1+e^{2\pi i p N }\,\overline{Y}(\theta)\big)\,
\big(1+e^{-2\pi i p N}\,\overline{Y}(\theta)\big)\, ,\cr 
&\overline{Y}\Big(\theta + {i\pi\xi\over2}\Big)\,
\overline{Y}\Big(\theta - {i\pi\xi\over2}\Big)=
\big(1+Y_{{N\over2}-1}(\theta)\big)\ ,}}
where as before $p$ denotes the eigenvalue of the operator ${\bf P}$. 
Note that the system  \DNtba\  coincides with the functional form of the
TBA equations of $D_N$ type \Zams, \FLS.

Further simplifications occur when 
$p$ takes special values. Suppose
that the operators 
${\bf T}_j$ act on the space ${\cal V}_p$ with
\eqn\pmn{p={\ell+1\over {2\, N}}\ ,}
where $\ell\geq 0$ 
is an integer such that $2p\not= n\beta^2 +m$ for any 
integers $n$ and $m$. Then  one can
show \ \BLZZZ\
that in addition to  \Tred\  the operators ${\bf T}_j$ satisfy
\eqn\Tredm{\eqalign{{\bf T}_{{N\over2}-j-1}(\lambda) 
= (-1)^{\ell}&\  {\bf T}_j (\lambda), \quad
\hbox{for} \quad \textstyle{j=0, \half , 1, ..., {N\over2}-1};\cr
&{\bf T}_{{N\over2}-\half }(\lambda) =
0\ .}}
This is exactly the case for
the ``minimal CFT'' ${\cal M}_{2/2n+3}$
with $n=1, 2, ...,$ where 
$$q=e^{{2\pi i}\over {2n+3}}$$
and the space of states has the form 
\eqn\Hmin{{\cal H}_{{\cal M}_{2/2n+3}} =
\oplus_{k=0}^{n}\  {\cal H}_{p_k}\ ,}
with
\eqn\pkmin{p_k = {{2k+1}\over {2(2n+3)}},
\qquad
\Delta(p_k) =-{(n+k+1)(n-k)
\over {2(2 n+3)}}; \ \ k=0,...,n  \ .} 
In this case the extra relations 
\Tredm\  allow one to bring \DNtba\  to
yet simpler form
\eqn\Mntba{\eqalign{
Y_j (\theta + i\pi/(2 n+1))\,Y_j (\theta - i\pi/(2 n+1)) &=
\big(1+Y_{j-\half }(\theta)\big)\,\big(1+Y_{j+\half }(\theta)\big),
\cr 
Y_0(\theta)=Y_{n+{1\over2}}(\theta)\equiv0;\qquad 
Y_{n+{1\over2}-j}(\theta) &= Y_j (\theta),\  j= \half , ..., n\ .}}
Again, \Mntba\  coincides with the 
functional form of the TBA equations
for the perturbed ``minimal CFT'' ${\cal M}_{2/2n+3}$ \ 
\Klassen. The above
truncation was discussed in \BLZ.

In this paper we concentrate attention on the simplest of the above
``minimal CFT'', the model ${\cal M}_{2/5}$. This CFT describes the 
criticality associated with the Lee-Yang edge
singularity \ \Cardyu\  and
therefore it is often
referred to as the Lee-Yang CFT. The central charge
is 
\eqn\cdd{c ({\cal M}_{2/5})= -{22\over 5}\ }
and the space \Hmin\  contains only two components
\eqn\Hdp{{\cal H}_{{\cal M}_{2/5}} =
{\cal H}_{p_0} \oplus {\cal H}_{p_1}\ ,}
where $p_0 = 1/10$ and $p_1 =3/10$,
the associated conformal dimensions
are
\eqn\Deltas{\Delta (p_0) =-1/5  , \qquad \Delta(p_1) = 0\  .}
For this model the functional equation \Mntba\  takes the form
\eqn\yyy{Y(\theta + i\pi/3)\, Y(\theta -i\pi/3) = 1 + Y(\theta)\ ,}
where according to \Yjdefa\  the function $Y(\theta)\equiv Y_\half(\theta)=
Y_1(\theta)$ simply coincides
with corresponding eigenvalue $T_{1/2}(\lambda)$
of the operator ${\bf T}(\lambda)
= {\bf T}_{1/2} (\lambda)$ in the space \Hdp,
\eqn\YT{Y(\theta) = 
T_{1/2}(\lambda)\ ,\ \ \  e^{\theta}=\lambda^{{5\over 3}}\
 R/(2 \pi)\ .}

The analytic properties of
the eigenvalues $T_{1/2}(\lambda)$ was discussed in
Sect.2.  They are entire functions of
the variable
$$t\equiv \lambda^2 \, \Big(\, {R\over 2\pi}\, \Big)^{{6\over 5}}$$
with the asymptotic behavior
\eqn\tass{\log T_{1/2}(\lambda) 
\sim \kappa\  t^{5\over 6}, \quad {\rm as}
\quad t\to \infty\  ,}
where
$$\kappa={2\sqrt{\pi}\, \Gamma(1/6)\over \Gamma(2/3)}\ \Big(\, 
\Gamma(3/5)\, \Big)^{{5\over 3}}= 28.29877111...\ , $$
as it follows from \coeffcn\ with \ $\beta^2=2/5$.
Correspondingly, we are interested here in the solutions $Y(\theta)$ of 
\yyy\  which are regular everywhere in the complex $\theta$-plane,
satisfy the periodicity condition \foot{In fact, this
periodicity condition can be derived from  \yyy,
 see \ \Zams.}
\eqn\yperiod{Y\big(\theta+{5\over 3}\, i\pi\big) = Y(\theta)\ ,}
and have the asymptotic
\eqn\yass{Y(\theta) \sim \exp(\kappa\,  e^{\theta}),
\quad {\rm as} \quad \theta \to +\infty\ .}
Note that since the eigenvalues $T_{1/2}(\lambda)$
are regular at $t=0$ the
associated functions $Y(\theta)$ are bounded at $\theta\to -\infty$;
in fact in this limit they approach the constant values
\eqn\yconstants
{Y(\theta) \to \cases{2\cos(2\pi p_0) = {(1+\sqrt{5})/2}\quad
\hbox{for the eigenvalues in} \quad {\cal H}_{p_0}\ ;\cr
2\cos(2\pi p_1) = {(1-\sqrt{5})/2}\quad
\hbox{for the eigenvalues in} \quad {\cal H}_{p_1}\ .\cr} }
as it follows from \eigenexp{a}.

The functional equation identical to \yyy\  
arises in the solvable lattice hard hexagon model \BaxPea.
It was studied by Klumper and Pearce\  \KluP, \ 
who developed the method of transforming the functional equations to
nonlinear integral equations, the knowledge about analytic properties
of the solutions being key ingredient in this approach. Although in
the field theory the analytic properties of the functions $Y(\theta)$ are
somewhat different from those considered in \KluP,
the method of 
Kl\"umper and Pearce can be easily
adopted to our case. The derivation 
in \ \KluP\  is based on the following

\vskip 0.1in
 
{\bf Lemma}. {\it If a function $f(\theta)$ is regular and bounded
in the strip $\Im m\, \theta \in (-\pi/3, \pi/3)$ and satisfies the 
relation
\eqn\lemmadif{f(\theta + i\pi/3) +
f(\theta - i\pi/3) - f(\theta) = g(\theta)\ }
with some $g(\theta)$ then
\eqn\lemmaint{
f(\theta) = \int_{-\infty}^{\infty}  d\theta'\ 
\Phi(\theta - \theta')\, g(\theta')\ ,}
where
\eqn\lemmaphi{\Phi(\theta) =
{\sqrt{3}\over \pi}\ {{\sinh(2\theta)}\over
{\sinh(3\theta)}}\ .}

}

Note that 
\eqn\phis{\Phi(\theta) = {i\over 2\pi}
\partial_{\theta}\log\, S(\theta)\ ,}
where $S(\theta)$ is the S-matrix \Smatrix. 
If $Y(\theta)$ was free of zeroes in the strip
$\Im m\,  \theta \in (-\pi/3, \pi/3)$ then the function 
$f(\theta)=\log Y(\theta) - \kappa\,\exp(\theta)$
would satisfy 
the conditions of this lemma with $g(\theta)= \log\big(
1+Y^{-1}(\theta)\big)$, so the integral equation would follow 
immediately. However a typical eigenvalue $T(t,p)$ does have zeroes in
the wedge $-2\pi/5 < \arg t < 2\pi/5$ (which corresponds to the above
strip) and so we need to take care of them.

Let us assume that $Y(\theta)$ has $N+2M$ zeroes in the strip 
$\Im m\,  \theta \in (-\pi/3, \pi/3)$, $N$ real zeroes $\alpha_a, a=
1,2,...,N$ and $M$ complex-conjugated pairs $(\beta_b + i \gamma_b,
\beta_b - i\gamma_b), b=1,2,...,M$. We introduce the following 
functions
\eqn\sigmas{\eqalign{&\ \ \ \ \ \ \sigma_0 (\theta) = \tanh\big({3\over
4}\, \theta\big)\ ,\cr
\sigma_1 (\theta, \eta) = &{{\cosh(\theta) - \cos(\eta)}\over
{\cosh(\theta) + \cos(\eta)}}\ {{\cosh(\theta) - \sin(\pi/6-\eta)}
\over {\cosh(\theta) + \sin(\pi/6 - \eta)}}\ ,}}
which satisfy the equations
\eqn\eqsigma{\sigma_0 (\theta + i\pi/3)\,
\sigma_0 (\theta - i\pi/3) = 1;
\qquad \sigma_1 (\theta + i\pi/3,\eta)
\, \sigma_1 (\theta - i\pi/3, \eta) = \sigma_1 (\theta, \eta)\ ,}
and write $Y(\theta)$ as
\eqn\Xdef{\eqalign{Y(\theta) 
&= \prod_{a=1}^{N}\sigma_0 (\theta - \alpha_a)\ 
e^{\epsilon(\theta)}\cr
&=\exp(\kappa\ e^\theta)\  
\prod_{a=1}^{N}\sigma_0 (\theta - \alpha_a)
\prod_{b=1}^{M}\sigma_1(\theta - \beta_b, \gamma_b) \ X (\theta)\ .}} 
In view of  \eqsigma\  the function $X(\theta)$ satisfies the equation
\eqn\Xeq{X(\theta+i\pi/3)\,  X^{-1}(\theta)\, X(\theta-i\pi/3)=
\prod_{a=1}^{N}\sigma_0 (\theta - \alpha_a)\big(1 +
Y^{-1}(\theta)\big)\ .}
Note that because the $\sigma$-factors in \Xdef\  absorb all zeroes
of $Y(\theta)$ in the strip  
$\Im m\  \theta \in (-\pi/3, \pi/3)$ and the
exponential prefactor saturates the asymptotic \yass\  the function 
$\log X(\theta)$ is regular and bounded in this strip. Applying the 
above lemma we obtain the integral equation
\eqn\TBA{\epsilon(\theta) = \kappa\, e^\theta + 
\sum_{b=1}^{M}\log\big(\sigma_1 (\theta - \beta_b, \gamma_b)\big)
+\int_{-\infty}^{\infty} d\theta'\ \Phi(\theta-\theta')\, 
\log \Big(\prod_{a=1}^{N}
\sigma_0 (\theta' - \alpha_a) + e^{-\epsilon(\theta')}\Big)\ ,} 
where $\epsilon(\theta)$ is defined in \Xdef\ \KluP. 

Given a set of real 
parameters $\alpha_a, \beta_b, \gamma_b$ the
equation \TBA\  uniquely
determines the function $\epsilon(\theta)$ and hence $Y(\theta)$. 
Of course these parameters can not be taken arbitrary. With generic
values of these parameters the function $Y(\theta)$ defined through
the solution to \TBA\  
will have the poles in the domain $\Im m\, \theta
\in (\pi/3, 2\pi)\ {\rm mod}\ 5\pi/3$. Indeed, if, for instance, 
$Y(\beta_b + i\gamma_b)
= 0$ then according to \yyy\  the function $Y(\theta)$ must exhibit a 
pole at $\theta = \beta_b + i\gamma_b - 2\pi i/3$, unless, of course, 
$Y(\beta_b + i\gamma_b - i\pi/3) = -1$. Therefore the condition that 
$Y(\theta)$ is entire function leads to the equations
\eqn\BAdress{\eqalign{
&Y(\alpha_a + i\pi/3) = Y(\alpha_a - i\pi/3) = -1,
\ \ \ a=1,...,N\  ;\cr
&Y(\beta_b + i\gamma_b - i\pi/3) = Y(\beta_b - i\gamma_b + i\pi/3) =
-1,\ \ \ b=1,...,M\ .}}
which determine the parameters $\alpha_a, \beta_b$ and $\gamma_b$ in
\Xdef.
The integral
equation \TBA\  together with the transcendental equations \BAdress\ are QFT
versions of the equations obtained in the lattice theory in \KluP\
 and therefore
we will refer to them as Kl\"umper-Pearce equations.

The Kl\"umper-Pearce equations \TBA, \BAdress\ 
 enables one to determine (in principle) any eigenvalue
$Y(\theta)$ of the operator ${\bf T}(\lambda)$ in the space \Hdp.
In this paper we consider only two simplest eigenvalues - the vacuum
eigenvalues $T(t,p_0)$ and
$T(t, p_{1})$ which correspond
to the vacuum vectors $|p_{0}>$ and $|p_{1}>$ in the spaces 
${\cal V}_{p_0}$ and ${\cal V}_{p_1}$ in \Hdp, respectively.
We will use the notations  $Y_0(\theta) = T
(t,p_0)$\ and $Y_1(\theta) = T(t,p_1)$
(which are not to be 
confused with the $Y_j$\  used in  $\Yjdefa - \Mntba$)
with $t = \exp(6\, \theta/5)$.
\vskip 0.1in
I. $\ Y_0(\theta)$. This eigenvalue corresponds to the ground
state in the space \Hdp. As discussed in Sect.2 all zeroes of 
$T(t,1/10)$ lie on the negative part of the real $t$-axis and so
all zeroes of $Y_0(\theta)$ are located along the lines $\Im m
\, \theta = 5\pi/6\,
(  {\rm mod} \ 5\pi/3)$.
In particular, there are no 
zeroes in the strip $\Im m\,  \theta \in (-\pi/3,\pi/3)$. Therefore in 
this case the equation \TBA\  takes the form
\eqn\tba{\epsilon_0(\theta) = 
\kappa\ \exp(\theta) + 
\int_{-\infty}^{\infty}d\theta'\ \Phi(\theta -
\theta')\, \log(1+e^{-\epsilon_0(\theta')})\ ,} 
where $\epsilon_0(\theta) = \log Y_0(\theta)$, and the equations 
\BAdress\  do not appear. The Eq.\tba\  is exactly the massless TBA 
equation associated with the Lee-Yang CFT\ \Zar.
The solution of this
equation was referred to as
the ``kink solution'' in \ \Zar\ 
where it was studied numerically; the
solution is plotted in Fig.1. 

\midinsert

\centerline{\epsfbox{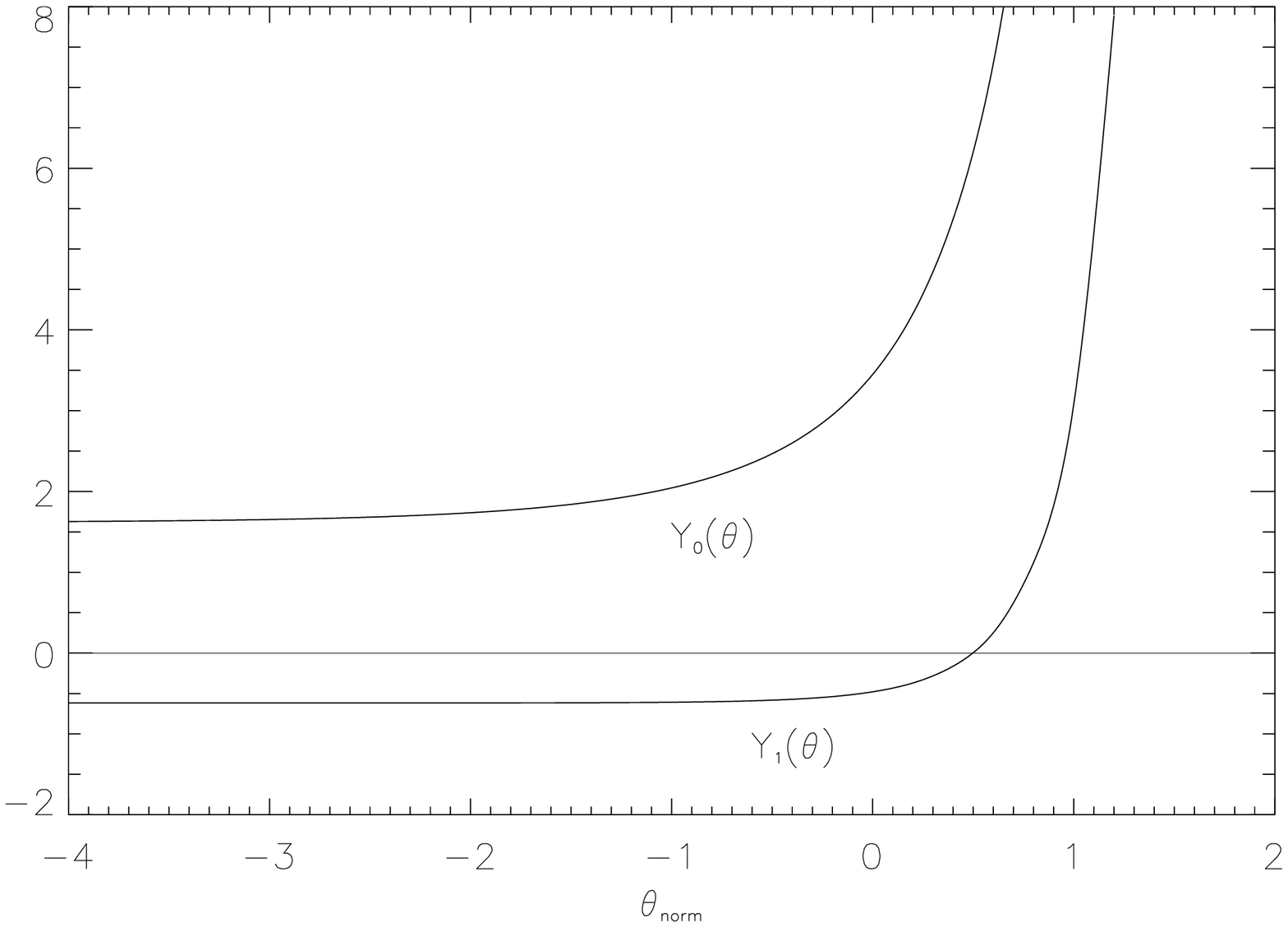}}
\capt{Fig.1. The functions  $Y_0(\theta)$ and $Y_1(\theta)$ found from 
numerical solutions of the integral equations \tba\ and (4.37)
\ plotted versus
a ``normalized'' 
rapidity variable $\theta_{{\rm norm}}=\theta+\log \kappa$.}
\endinsert

Note that for $\theta \to -\infty$
the function $Y_0(\theta)$ approaches the value
$2\cos(\pi p_{0}) = (\sqrt{5}+1)/2=1.6180339\ldots$ in
agreement with  \yconstants. We have used
the numerical solution to the equation \ \tba\  to evaluate few first 
coefficients in the expansion \eigenexp{a}; the result is given in Table~1 
where the numerical results are also compared with the exact 
coefficients calculated as the integrals \intgk\ (see Appendix A). 

\midinsert
\bigskip
\centerline{
\noindent\vbox{\offinterlineskip
\hrule
\halign{\vrule#&
  \strut\quad#\hfil\quad& 
        \vrule#& 
  \strut\quad\hfil#\quad&
        \vrule#\cr
height4pt&\omit&&\omit&\cr
&\hfil Exact values && Numerical values\hfil&\cr
height4pt&\omit&&\omit&\cr
\noalign{\hrule}
height4pt&\omit&&\omit&\cr
&$G^{vac}_0({1\over10})   
\gap=1.6180339887\ldots\,\phantom{\, 10^{1}}$
&&$1.618033989\ldots \phantom{\, 10^{1}}$&\cr
height4pt&\omit&&\omit&\cr
&$G^{vac}_1({1\over10})
\gap=7.01811993\ldots\,  10^{1}$
&&$7.0181198\ldots \,10^{1}$&\cr
height4pt&\omit&&\omit&\cr
&$G^{vac}_2({1\over10})
\gap=1.361347\,\ldots
10^{3}$
&&$1.36135\ldots \,10^{3} $&\cr
height4pt&\omit&&\omit&\cr
&$G^{vac}_3({1\over10}) \gap=
1.6320\ldots\, 10^{4}$
&&$1.6317\ldots \, 10^{4}$&\cr
height4pt&\omit&&\omit&\cr
&$G^{vac}_4({1\over10})\gap=$(unknown)
&&$1.41\ldots \,10^{5}$&\cr
height4pt&\omit&&\omit&\cr}
\hrule}
}

\bigskip
\capt{Table 1. The exact eigenvalues 
of a few first NIM for the $\Delta=-{1\over5}$\     vaccum state 
in the Lee-Yang CFT (as given by 
(A.2) and (A.9)  with $a=2/5$ and $p=1/10$) and numerical values 
of the same eigenvalues obtained with a polynomial fit of $Y_0(\theta)$
determined from     the numerical solution of the 
integral equation \tba. The quantity $G^{vac}_0(p)$ denotes the constant term 
in the series expansion \eigenexp{a}}
\endinsert

As is explained in 
Sect.2, the function $\epsilon_0(\theta) - \kappa\exp(\theta)$ can be 
expanded into asymptotic series in powers of $\exp(-\theta)$. The
easiest way to obtain this expansion is to note that the kernel
$\Phi$ in \tba\  expands for $\Re e\,  \theta > 0$ as
\eqn\phiexp{
\Phi(\theta) = -{2\over \pi}\
\sum_{n=1}^{\infty}\sin{{2\pi(n+1)}\over
3}\  e^{(1-2n)\theta}\ .} 
Substituting this into \tba\ one finds
\eqn\epsexp{\log Y_0(\theta) \simeq \kappa\,  e^\theta -
{2\over \pi}\, 
\sum_{n=1}^{\infty} 
\sin{{2\pi(n+1)}\over 3}\  
\happa_n \  e^{(1-2n)\theta}\ ,} 
where the coefficients $\happa_n$ are calculated as
\eqn\kapaint{\happa_n =
 \int_{-\infty}^{\infty}d\theta\ e^{(2n-1)\theta}\ 
\log(1+e^{-\epsilon_0(\theta)})\ .}
According to \eigenexp{b}\   the coefficients in \epsexp\ 
are related to the 
vacuum eigenvalues of the local IM ${\bf I}_{2n-1}$,
\eqn\kapalim{ I^{vac}_{2n-1}(-1/5)=
 {2\over \pi C_n}\  \sin{{2\pi(n+1)}\over 3 }\   \happa_n\ ,}
where 
\eqn\cnly{C_n = {5\over 3}{\sqrt{\pi}\over {n!}}\ 
\Big({2\over 5}\Big)^n\  
{{\Gamma\big({10 n-5\over 6}\big)}
\over {\Gamma\big({{2 n+2}\over 3}\big)}}\ 
\Big(\Gamma(3/5)\Big)^{{5-10 n\over 3}}\ .}
The first of the integrals   \kapaint, can be evaluated 
exactly using the well known ``dilogarithm trick''\  
(see  Appendix C)
\eqn\kapaa{\happa_1 ={\pi^2\over 15}\
\kappa^{-1}\ .}
This agrees with the value $I^{vac}_1(-1/5) = -1/60$. We used the 
numerical solution to \tba\  to evaluate few further
coefficients \kapaint; these results are compared with the exact 
eigenvalues $I^{vac}_{2n-1}(-1/5)$ in Table~2.

\midinsert
\bigskip
\centerline{
\noindent\vbox{\offinterlineskip
\hrule
\halign{\vrule#&
  \strut\quad#\hfil\quad& 
        \vrule#& 
  \strut\quad\hfil#\quad&
        \vrule#\cr
height4pt&\omit&&\omit&\cr
&\hfil Exact values && Numerical values\hfil&\cr
height4pt&\omit&&\omit&\cr
\noalign{\hrule}
height4pt&\omit&&\omit&\cr
&$I_1(-{1\over5})=-{1\over 60}\gap=-1.6666666666666\ldots\, 10^{-2}$
&&$-1.6666666666665\ldots \, 10^{-2}$&\cr
height4pt&\omit&&\omit&\cr
&$I_5(-{1\over5})={89\over756000}\gap=\fanm 1.17724867724\ldots\,  10^{-4}$
&&$\fanm1.1772486773\ldots \,10^{-4}$&\cr
height4pt&\omit&&\omit&\cr
&$I_7(-{1\over5})= - {211\over8100000}\gap=-2.6049382716\,\ldots
10^{-5}$
&&$-2.604938279\ldots \,10^{-5} $&\cr
height4pt&\omit&&\omit&\cr
&$I_{11}(-{1\over5})={  2160997 \over 464373000000}\gap=
 \fanm  4.6535802\ldots\, 10^{-6}$
&&$\fanm 4.653584\ldots \, 10^{-6}$&\cr
height4pt&\omit&&\omit&\cr
&$I_{13}(-{1\over5})=-{ 6283403\over 1924560000000}
\gap=  -3.264851\ldots\, 10 ^{-6}$
&&$   -3.26488\ldots  \,10^{-6}$&\cr
height4pt&\omit&&\omit&\cr
&$I_{17}(-{1\over5})=${ (unknown)}
&&$\fanm3.5501\ldots \,10^{-6}$&\cr
height4pt&\omit&&\omit&\cr}
\hrule}
}

\bigskip
\capt{Table 2. The exact vacuum eigenvalues 
of a few first (non-vanishing) LIM given by 
(B.1)-(B.8) with $c=-22/5$ and $\Delta=-1/5$ and numerical values 
for the same LIM
 obtained from \kapalim\ with the numerical solution of the 
integral equation \tba.}
\endinsert

\vskip 0.1in

II. $\ Y_1(\theta)$. According to our discussion in Sect.2 all zeroes
of the function $T(t,3/10)$ are real, just one of them being positive. 
Correspondingly, the function $Y_1(\theta)=T(t,3/10)$ 
has one real zero which we denote 
$\alpha$ and infinitely many zeroes located along the lines $\Im m\, 
\theta = 5\pi/6\  ({\rm mod} \ 5\pi/3)$. Therefore in this case the 
function $\epsilon_1(\theta)=\log Y_1(\theta)-\log\sigma_0 (\theta -
\alpha)$ solves the equation
\eqn\tbaa{\epsilon_1(\theta) = \kappa\ e^\theta + \int_{-\infty}^{\infty}
d\theta'\ \Phi(\theta - \theta')\, \log\Big({\rm tanh}
\big(3(\theta' - \alpha)/4
\big) + 
e^{-\epsilon_1(\theta')}\Big)\ ,}
and the parameter $\alpha$ satisfies
\eqn\baa{i\ e^{\epsilon_1(\alpha + i\pi/3)} =
-i \ e^{\epsilon_1(\alpha - 
i\pi/3)} = -1\ .}
The last relation can be written as
\eqn\baaa{\eqalign{\sqrt{3}\, \kappa\ e^{\alpha} +
{3\over {\pi}}\, &\vpint
d\theta\ 
{{\cosh\big(2(\theta-\alpha)\big)}
\over {\sinh(3\big(\theta-\alpha)\big)}}\ \log
\Big(\tanh\big(3(\theta - \alpha)/4\big) 
+ e^{-\epsilon_1(\theta)}\Big)=\ \cr
&=\pi (1+4 N)\ ,}}
where $N$ is some integer and ${\rm -}\kern -.85em\int_{-\infty}^\infty$ 
denotes the principal value
of the singular integral.
We have used \tbaa\  and
explicit form of the kernel $\Phi(\theta)$\ \lemmaphi\ 
to obtain \baaa.
We studied the equations
\tbaa\ and \baaa\  numerically.  Our analysis suggests that these 
equations admit real solution only if $N$ in \baaa\ is zero, in which case
\eqn\thetaa{\alpha+\log \kappa = 0.49577315\ldots\ . }
The corresponding
function
\eqn\Yone{Y_1(\theta)=\sigma_0(\theta-\alpha)\, e^{\epsilon_1(\theta)}}
is plotted in Fig.1.
In this case the asymptotic value of $Y_1(\theta)$ at $\theta \to
-\infty$ is $2\cos(2\pi p_{1})
= (1-\sqrt{5})/2=-0.61803398\ldots$. 
Few first coefficients 
of the expansion \eigenexp{a}\  calculated from this numerical
solution together with their exact values (when available) are presented
in Table~3. 

\midinsert
\centerline{
\noindent\vbox{\offinterlineskip
\hrule
\halign{\vrule#&
  \strut\quad#\hfil\quad& 
        \vrule#& 
  \strut\quad\hfil#\quad&
        \vrule#\cr
height4pt&\omit&&\omit&\cr
&\hfil Exact values && Numerical values\hfil&\cr
height4pt&\omit&&\omit&\cr
\noalign{\hrule}
height4pt&\omit&&\omit&\cr
&$G^{vac}_0({3\over10})=\ \ 
\gap-0.6180339887\ldots\,$
 &&$-0.618033989\ldots \phantom{\, 10^{1}}$&\cr
height4pt&\omit&&\omit&\cr
&$G^{vac}_1({3\over10})=\ \ \ 0$
&&$3.6\ldots \,10^{-5}$&\cr
height4pt&\omit&&\omit&\cr
&$G^{vac}_2({3\over10})=
\gap2.94659\,\ldots 10^{2}$
 &&$2.9467\ldots \,10^{2} $&\cr
height4pt&\omit&&\omit&\cr
&$G^{vac}_3({3\over10}) = \gap
5.80714\ldots\, 10^{3}$
&&$5.80705\ldots \, 10^{3}$&\cr
height4pt&\omit&&\omit&\cr
&$G^{vac}_4({3\over10})=\gap6.2826\ldots\,10^{4}$
&&$6.281\ldots \,10^{4}$&\cr
height4pt&\omit&&\omit&\cr
&$G^{vac}_5({3\over10})=\gap4.729\ldots\,10^{5}$
&&$4.75\ldots \,10^{5}$&\cr
height4pt&\omit&&\omit&\cr
&$G^{vac}_6({3\over10})=\gap2.72\ldots\,10^{6}$
&&$2.6\ldots  \,10^{6}$&\cr
height4pt&\omit&&\omit&\cr
&$G^{vac}_7({3\over10})=\gap$(unknown)
&&$1.7\ldots \,10^{7}$&\cr
height4pt&\omit&&\omit&\cr}
\hrule}
}

\bigskip
\capt{Table 3. The exact eigenvalues 
of a few first NIM for the $\Delta=0$ vacuum state 
in the Lee-Yang CFT (as given by 
(A.2), (A.7), (A.8), (A.10)
with $a=2/5$ and $p=3/10$) and numerical values 
of the same NIM obtained with a polynomial fit of $Y_1(\theta)$
determined from the numerical solution of the 
integral equation \tbaa.
The quantity $G^{vac}_0(p)$ denotes the constant term 
in power series expansion \eigenexp{a}}

\endinsert

To obtain the asymptotic expansion for $\log 
Y_1(\theta)$ similar to \epsexp\  one could try again to substitute
the expansion \phiexp\  into \tbaa. However in this case the $\log$
factor in the integrand in  \tbaa\  does not decay sufficiently fast 
as $\theta' \to \infty$ and so the integrals appearing after this
substitution typically diverge. To handle this problem it is convenient
to rewrite the equation \tbaa\  as
\eqn\tbaaa{\eqalign{
\log|\, Y_1(\theta)\,| =
\kappa\ e^\theta& - {1\over 2}\, \log\big(S(\theta - \alpha
+ i\pi/3)\, S(\theta - \alpha + 2i\pi/3)\big) +\cr 
&\int_{-\infty}^{\infty}d\theta'\ \Phi(\theta - \theta')\,
\log\big(\mid 1 + Y^{-1}_1(\theta')\mid \big)\ .}}
Now using \phiexp\  we obtain the asymptotic
series expansion
\eqn\assexp{\log Y_1(\theta) = \kappa\ e^\theta - 
{2\over \pi}\ \sum_{n=1}^{\infty}\,  \sin{{2\pi(n+1)}\over 3}\ 
{\tilde \happa}_n\  e^{(1-2n)\theta}\ ,}
where
\eqn\kappaint{{\tilde \happa}_n ={2\pi\, \over  2 n-1}\
\sin{2\pi ( n+1)\over 3}\ 
e^{\alpha (2 n-1)} + \int_{-\infty}^{\infty}d\theta\ e^{(2n-1)\theta}
\log\big(\mid 1 + Y^{-1}_1(\theta)\mid \big)\ .}
\midinsert
\centerline{
\noindent\vbox{\offinterlineskip
\hrule
\halign{\vrule#&
  \strut\quad#\quad& 
        \vrule#& 
  \strut\quad\hfil#\quad&
        \vrule#\cr
height4pt&\omit&&\omit&\cr
&\hfil Exact values\hfil && Numerical values\hfil&\cr
height4pt&\omit&&\omit&\cr
\noalign{\hrule}
height4pt&\omit&&\omit&\cr
&$I_1(0)={11\over60}\gap=\fanm1.833333\ldots\,10^{-1} 
$&&$1.83332\ldots \,10^{-1}$&\cr height4pt&\omit&&\omit&\cr
&$I_5(0)={341\over756000}\gap
=\,\,\fanm4.51058\ldots\,10^{-4}$&&$4.5106\ldots \, 10^{-4}
$&\cr height4pt&\omit&&\omit&\cr
&$I_7(0)=-{451\over8100000}\gap\gap
=\,- 5.567901\ldots\,10^{-5}$&&$-5.56789\ldots \,10^{-5}
$&\cr height4pt&\omit&&\omit&\cr
&$I_{11}(0)={3373117\over464373000000}\,\,\,\,\,\,\gap
=\fanm7.26380 \ldots\,10^{-6}  
$&&$7.2631\ldots \,10^{-6}$&\cr height4pt&\omit&&\omit&\cr
&$I_{13}(0)=-{825517\over174960000000}\gap=-4.7183\ldots\,10^{-6}  $&&$
-4.717\ldots \,10^{-6}$&\cr height4pt&\omit&&\omit&\cr
&$I_{17}(0)=${ (unknown)}\hfill
&&$\,  4.67\ldots  \,10^{-6}$&\cr
height4pt&\omit&&\omit&\cr}
\hrule}
}

\bigskip
\capt{Table 4. The exact vacuum eigenvalues 
of a few first (non-vanishing) LIM given by 
(B.1)-(B.8) with $c=-22/5$ and $\Delta=0$ and numerical values 
for the same LIM
 obtained from  (4.46)  with the numerical solution of the 
integral equation \tbaa.}
\endinsert
Like in the previous example, for $n=1$ the integral in \kappaint\ 
can
be calculated analytically using appropriate modification of the
``dilogarithm trick'' (see Appendix C),
\eqn\tkappaa{{\tilde\happa_1} =-{11\,  \pi^2\over 15}\ \kappa^{-1}\ . }
The vacuum eigenvalues $I^{vac}_{2n-1}(0 )$ of the local IM are expressed 
through ${\tilde \happa}_n$ as
\eqn\kappalim{I^{vac}_{2n-1}(0)=
{2\over \pi C_n}\, \sin{{2\pi(n+1)}\over 3}\ {\tilde \happa}_n\ ,}
where $C_n$ are given by  \cnly, so that \tkappaa\  agrees exactly with
$I^{vac}_1 (0) = 11/60$. The values of few further local IM obtained from
the above numerical solution are compared with the exact values in
Table~4.

\newsec{Excited States  in Scaling Lee-Yang Model}

In this section we apply the approach outlined in Sect.4 to the finite
size energy spectrum of the SLYM \SLYM. The QFT \SLYM\ is particular 
case of \action\  and so 
it possesses infinitely many local IM \pertloim. 
In fact, due to the strong degeneracy of the representations entering
\Hdp\  in this case some of these IM vanish. Only the IM ${\Bbb
I}_{2n-1}$ with $2n-1 \neq 0 \, ({\rm mod}\, 3)$
are nonzero. As is mentioned
in the Introduction the finite-size ground state energy $E_0 (R)$ of 
this model was studied by TBA in \Zar.
In this section we show that the 
approach based on the operators \fullT\  and the fusion relations 
\fullfusion\  makes it possible to study the excited state energies in 
the finite size system as well. 

As was explained above, the perturbed theory inherits the space of
states from the CFT ${\cal M}_{2/5}$, that is it has the form
\eqn\hly{{\cal H}_{SLYM} = {\cal H}_{p_{0}}\,
\oplus {\cal H}_{p_{1}}\ ,} 
where $ p_{0} =1/10$ and $ p_{1} = 3/10$.
Consider the operator ${\Bbb T}(\mu|\lambda) \equiv {\Bbb T}_{1/2}
(\mu|\lambda)$, as defined in \fullT. This operator invariantly act in the 
space \hly\  and when specialized to this space it satisfies the
truncated fusion relation
\eqn\TTT{{\Bbb T}(\mu |q^{1\over 2}\lambda)\,
{\Bbb T} (\mu| q^{-{1\over 2}}\lambda)
= {\Bbb I} + {\Bbb T}(\mu|\lambda)\ ,}
and therefore the eigenvalues $T(\mu | \lambda)$
of ${\Bbb T}(\mu | \lambda)$
in the space \hly\  satisfy exactly the same functional equation 
\eqn\ttt{T(\mu|q^{1\over 2}\lambda)\,
T(\mu|q^{-{1\over 2}}\lambda) = 1+T(\mu|\lambda)}
as the eigenvalues of the operator ${\bf T}$ in the space \Hdp.
However, in the perturbed case ($\mu \neq 0$) we are interested in the
solutions to \ttt\  which have significantly different analytic 
properties as the functions of $\lambda^2$. As was explained in
Sect.3, the eigenvalues
$T(\mu | \lambda)$ have essential singularities both
at $\lambda^2 = \infty$ and $\lambda^2 = 0$,
and they exhibit the asymptotic behavior
dictated by \fullass\  and \fullasss\  near these singularities.

Like in the Sect.4 it is convenient here to introduce the rapidity
\eqn\thetat{\theta = {5\over 6}\  \log\big(\, \lambda^2/\mu\, \big)\ . }
Note that this definition differs from  \YT\  by a constant
shift $\log(\, \mu^{5\over 6}\,R/2\pi\,$).
Also, we will use the dimensionless 
variable
\eqn\rR{ r= m\, R \equiv \big(\, \kappa\,\mu^{5\over 6}/\pi\,\big)\, R\ }
instead of $R$. Here $m$ is the mass related to the coupling parameter
${\hat \mu}$\ \SLYM\  as in \mumly. 
The eigenvalues $T(\mu|\lambda)$ can be considered as the
functions of $\theta$ and $r$. Like in \YT\  we denote
\eqn\YTT{Y(r |\theta) = T(\mu|\lambda)\, ;\  \qquad r = 
\kappa\,\mu^{5\over 6}\ 
R/\pi\, , \ \ e^{\theta}=\lambda^{5\over 3}\, 
\mu^{-{5\over 6}}
\ .}
In this notations the functional equation \ttt\  takes the form \yyy,
i.e.
\eqn\YYY{
Y(r|\theta+i\pi/3)\,Y(r|\theta-i\pi/3) = 1 + Y(r|\theta)\ .}
We are interested now
in the solutions which are analytic everywhere in the 
complex $\theta$-plane, satisfy the periodicity
condition \yperiod\  
and have the asymptotic
\eqn\Yass{\log Y(r|\theta)
\sim r \ e^{|\theta|}/2\ , \ \ \theta\to\pm \infty\ .} 

Like in the conformal case discussed in Sect.3 the solution to \YYY\  
with the asymptotic conditions \Yass\  is completely determined by the 
pattern of zeroes of $Y(r|\theta)$ in the strip 
$-\pi/3 < \Im m\, \theta < \pi/3$. The same transformations as
in Sect.3
lead to the integral equation
\eqn\TBAA{\eqalign{\epsilon(r|\theta) =& r\cosh\theta + 
\sum_{b=1}^{M}\log\sigma_1 \big(\theta-\beta_b, 
\gamma_b \big) +\cr
&\int_{-\infty}^{\infty}d\theta'\ \Phi(\theta-\theta')\log
\ \big(\prod_{a=1}^{N}\sigma_0 (\theta'-\alpha_a ) + 
e^{-\epsilon(r|\theta')} \big)\ ,}}
where
\eqn\ey{\epsilon(r|\theta) = \log Y(r|\theta) - \sum_{a=1}^{n}\log\sigma_0
(\theta - \alpha_a ) }
and we have assumed that there are $N+2M$ zeroes in the physical strip,
$N$ real zeroes $\alpha_a (r)$ and $M$ complex-conjugated pairs
$\beta_b (r) \pm i\, \gamma_b (r)$; here we have shown explicitly that the
positions of these zeroes depend on the parameter $r$\ \ \foot
{We will see later that when $r$ change some complex-conjugate zeroes
can collide and then become a pair of the real zeroes, or vice versa,
so that in different domains of $r$ the numbers $N$ and $M$ may vary,
while the total number of zeroes, of course, remains the same.}. 
And again, like in Sect.4 these positions are determined by
the equations
\eqn\BAfulla{Y(r|\alpha_a  + i\pi/3) =
Y(r|\alpha_a  - i\pi/3) = -1, \ \ a=1,...,N\  ;}
\eqn\BAfullb{Y(r|\beta_b + i\gamma_b - i\pi/3) =
 Y(r|\beta_b -i\gamma_b + i\pi/3)
= -1,\ \ b=1,...,M\ .}

Before studying particular examples of the eigenvalues $Y(r|\theta)$
let us show how the mass spectrum and factorizable S-matrix of this 
QFT can be recovered from these equations in the limit $r\to\infty$.
Simple estimate shows that for $r>>1$ the first term in the r.h.s. 
of Eq.\TBAA\  dominates everywhere in the strip $\Im m\, \theta \in 
(-\pi/3 + \varepsilon\,,\,\pi/3 - \varepsilon)$ except in the small disks 
of the
size $\varepsilon$ around the zeroes $\alpha_a $ and $\beta_b \pm
i\, \gamma_b$, where $\varepsilon$ is some number exponentially small in
$r$. In this domain the first term in the r.h.s of \YYY\  can be 
neglected and hence the function $Y(r|\theta)$ can be written as
\eqn\prodsigma{\eqalign{&Y(r|\theta) \simeq \prod_{a}
S(\theta-\beta_{a}-i\pi/2)
\exp(r\cosh\theta)\ ,\cr
&\qquad\ \ \ \ \ \  r>>1; \qquad |\theta -\beta_{a} \pm i\pi/6 | >
\varepsilon; \qquad \Im m\, \theta \in (-\pi/3 +\varepsilon,
\pi/3-\varepsilon)\  ,}}
where $S(\theta)$ is
the S-matrix \ defined by\ \Smatrix\     
and $\beta_a$ are some real numbers. In writing \prodsigma\  we
have taken into account the asymptotic conditions \Yass\  and the fact 
that $Y(r|\theta)$ must be real at real $\theta$ \foot{
The factors of the form $S(\theta +
i\pi/6)\, S(\theta+5i\pi/6)$ also satisfy \YYY\
with the first term in 
the r.h.s. in this equation omitted,
and they agree with the 
real analyticity condition. More detailed analysis shows however that
these factors would be 
incompatible  with \BAfulla,\ \BAfullb.}. The form
\prodsigma\  shows in particular that for $r$ sufficiently large there
are no real zeroes $\alpha_a$ in \TBAA\  while the complex zeroes have
the form
\eqn\nudef{\beta_b (r) \pm i \, \big(\pi/6 - \nu_b (r)\big)\ ,}
where $\beta_a (r)$ are real and $\nu_a (r)$ are exponentially 
small in $r$. The equation \BAfullb\  then takes the form
\eqn\BAnaive{\exp(i\,r\,\sinh\beta_b )\ \prod_{b\neq a}S\big(\beta_b  -
\beta_a \big) 
= 1, \quad b = 1,2,...,M\  ,}
where we have taken into account that $S(0)=-1$.
This equations shows that
for $r>>1$ the
eigenstate associated with the eigenvalue\ \prodsigma\  
is interpreted as M-particle state, the parameters $\beta_b =\beta_b (r)$ 
being
the rapidities of the particles,
so that the equations \BAnaive\  have the 
meaning of the Bethe Ansatz equations for $M$ scalar particles with the 
S-matrix \Smatrix \ \Zar\ . We should stress here that in the field theory the 
Bethe Ansatz equations of the form \BAnaive\  make sense only in the 
limit $r\to\infty$. For finite $r$ these equations acquire corrections 
due to the vacuum polarization.

Here we consider in details only two simplest eigenvalues of the
operator ${\Bbb T}$ of the Lee-Yang model.

\vskip 0.1in
  
I.  The simplest eigenvalue is of course the ground state eigenvalue,
i.e. the eigenvalue associated with the lowest eigenstate of the 
Hamiltonian \hamilton\  in the space \hly. We denote here this state 
as $\mid \Psi_0 \rangle$ and use the notation $Y_0(r|\theta)$ for the
corresponding eigenvalue of ${\Bbb T}$. For the ground state eigenvalue 
$Y_0(r|\theta)$ there are no zeroes in the  strip $\Im m\  
\theta \in (-\pi/3,\pi/3)$ at all and the equation \TBAA\  takes the form
\eqn\vactba{
\epsilon_0 (r|\theta) = 
r\cosh\theta + \int_{-\infty}^{\infty} d\theta'\ 
\Phi(\theta-\theta')\ \log(1+e^{-\epsilon_0 (r|\theta')})\ ,}
where $\epsilon_0 (r|\theta) = \log Y_0 (r|\theta)$. This is exactly
the TBA equation for the Lee-Yang model obtained in\ \Zar; here we have 
arrived
at \vactba\  by entirely different route, completely bypassing any
reference to the on-shell solution of the model.
The equation \vactba\  
was studied in very details in \ \Zar\
and we can add very little to that
analysis. Let us just briefly summarize
the results of \ \Zar. For large
$r$ the first term in
the r.h.s. of \vactba\  dominates and so 
$\epsilon_0 (r|\theta) \sim r\cosh\theta; \ \ r>>1$. In the opposite
limit  $r \to 0$ the function $Y_0 (r|\theta)$ becomes nearly constant
equal to ${\sqrt{5}+1\over 2}$\  in the wide range of $\theta$, 
$\log r < \theta < -\log r$, while for $\theta \gtrsim -\log r$ this 
function coincides with the ``kink solution'' $Y_0\big(\theta +
\log\big( {r\over 2\kappa}\big)\big)$
considered in Sect.4; similarly, 
for $\theta \lesssim \log r$,
\  $Y_0 (r|\theta)\to Y_0\big(-\theta-\log\big(
{r\over 2\kappa}\big)\big)$. In view of \zeromu\  and \ \zeromubar\ 
this is just the manifestation of the 
fact that for $r\to 0$ the ground
state $\mid \Psi_0 \rangle$ approaches
the conformal vacuum state $\mid \Delta(p_0) \rangle $\ \erw\ 
in \hly. According to
\fullTexp\
the ground state eigenvalues of the local IM \pertloim
\eqn\gsloim{\eqalign{&{\Bbb I}_{2n-1}\, 
\mid \Psi_0 \rangle= {I}^{(0)}_{2n-1}(R)\,
  \mid \Psi_0 \rangle\, \cr
&{\bar {\Bbb I}}_{2n-1}\,
\mid \Psi_0 \rangle= {\bar {I}}^{(0)}_{2n-1}(R)\,
  \mid \Psi_0 \rangle\ } }
can be obtained from $\epsilon_0 (r|\theta)$
as the coefficients of the
asymptotic expansion
\eqn\gdfd{\eqalign{
&\epsilon_0 (r|\theta) \simeq r\, e^\theta/2 - \sum_{n=1}^{\infty}
C_n\ {\mu}^{5(1-2n)/6}\ e^{(1-2n)\theta}\ I^{(0)}_{2n-1}(R)\, ,
\ \ \  \theta\to
+\infty\ ,\cr
&\epsilon_0 (r|\theta) \simeq r\, e^{-\theta}/2 - \sum_{n=1}^{\infty}
C_n\ {\mu}^{5(1-2 n)/6}\ e^{(2n-1)\theta}\ {\bar I}^{(0)}_{2n-1}(R)\, ,
\ \ \   \theta\to-\infty\ .}} 
It follows from \vactba\  that
\eqn\iko{\eqalign{
& I_{1}^{(0)}(R)
= -{\sqrt{3}\over 24}\, m^2 R -{m\over 4 \pi}\  
\int_{-\infty}^{\infty}d \theta\ e^\theta
\log\big(1+e^{-\epsilon_0 (r|\theta)}\big)\ ,\cr
&\ I_{2n-1}^{(0)}(R) = {2\over \pi C_n}\ \Big({\pi m\over
\kappa
}\Big)^{1-2 n }\  
\sin{2\pi (n+1)\over 3}\ 
\int_{-\infty}^{\infty} d\theta\ e^{(2n-1)\theta}
\ \log(1+e^{-\epsilon_0 (r|\theta)})\ ,}}
where\ $ n=2,3,...\ .$
The similar relations hold for the  ``right"
local IM  \ ${\bar I}_{2n-1}^{(0)}(R)$. Then, 
for the ground-state energy 
$$ {\Bbb H} \mid \Psi_0\rangle=E_0 (R)\mid \Psi_0\rangle\ ,$$
we  have
\eqn\eo{E_0 (R) \equiv I_{1}^{(0)}(R)+{\bar I}_{1}^{(0)}(R)=
-{\sqrt{3}\over 12}\, m^2 R -{m\over 2\pi}\ 
\int_{-\infty}^{\infty}d\theta\ \cosh\theta \ 
\log\big(1+e^{-\epsilon_0 (r|\theta)}\big)\ . }
Obviously it is the first term in \ \eo\  that dominates at 
large $R$ and therefore it gives the bulk vacuum energy of SLYM while
the second term represents the finite-size corrections.
Finally, the leading
$R\to 0$ asymptotic
\eqn\eouv{E_0 (R) \sim - {\pi\over 15\, R} }
can be deduced from\  \eo\  in full agreement with expected form 
$E_0(R) = -(\pi/6R)(c-24\Delta(p_0))$.

II.  Let us consider now the first excited state of the SLYM. We will
denote this state and associated eigenvalue of ${\Bbb T}$ as 
$\mid \Psi_1  \rangle$ and $Y_1 (r|\theta)$, respectively. At $R\to
\infty$ this state is interpreted as one-particle state with the
particle momentum (and rapidity) equal zero. Therefore, according to our
discussion above, for large $r$ the eigenvalue $Y_1 (r|\theta)$ has two 
zeroes in the physical strip located at 
\eqn\eozero{
\theta =  \pm \,i \,\gamma (r) = \pm\, i\, 
\big(\pi/6 - \nu (r)\big)\ ,}
where $\nu (r) \to 0$ as $r \to \infty$ with exponential accuracy.
Correspondingly, in this case the integral equation \TBAA\  for the 
function $\epsilon_1 (r|\theta)=\log  Y_1 (r|\theta)$ takes the form
\eqn\tbaee{\epsilon_1 (r|\theta) = r\cosh\theta + 
\log \sigma_1 \big(\theta ,
{\pi\over 6}-\nu \big) + \int_{-\infty}^{\infty}d\theta'\ 
\Phi(\theta - \theta')\  \log (1 + e^{-\epsilon_1 (r|\theta')})\ ,}
where  $\sigma_1(\theta,\eta)$\ is given  in \sigmas\ 
and the function $\nu (r)$ which describes
the deviation of zeroes from their limiting
positions at $\pm i\pi/6$  is determined from the equation \BAfullb  
\eqn\BAexcaa{e^{\epsilon_1 (r|\,\pm i(\pi/6+\nu))} = -1\ .}
In the limit $r >> 1$ the term with the integral in \tbaee\  is
exponentially small and with this accuracy one has
\eqn\yoass{Y_1 (r|\theta) = {{\cosh\theta-\cos({\pi\over 6}-\nu)}\over 
{\cosh\theta+\cos({\pi\over 6}-\nu)}}\ {{\cosh\theta -\sin\nu}\over
{\cosh\theta+\sin\nu}}\ e^{r \cosh\theta}\ \big(1 + O(e^{-r})\big)\ ,}
where $\nu = \nu (r)$ is derived from \BAexcaa
\eqn\nuass{
\nu (r) = \sqrt{3}\ e^{-{\sqrt{3}\over 2}\,  r} +
O\big(re^{-\sqrt{3}\, r}\big)\ .}

For $r \sim 1$ the equations \tbaee,\  \BAexcaa\  can be solved 
numerically. The resulting function $\gamma (r)$ is given in
Table~5 for $2.6 < r < 8.0$ and the typical form of the function
$Y_1 (r|\theta)$ in this range of $r$ is shown in Fig.2.
\midinsert

\centerline{\epsfbox{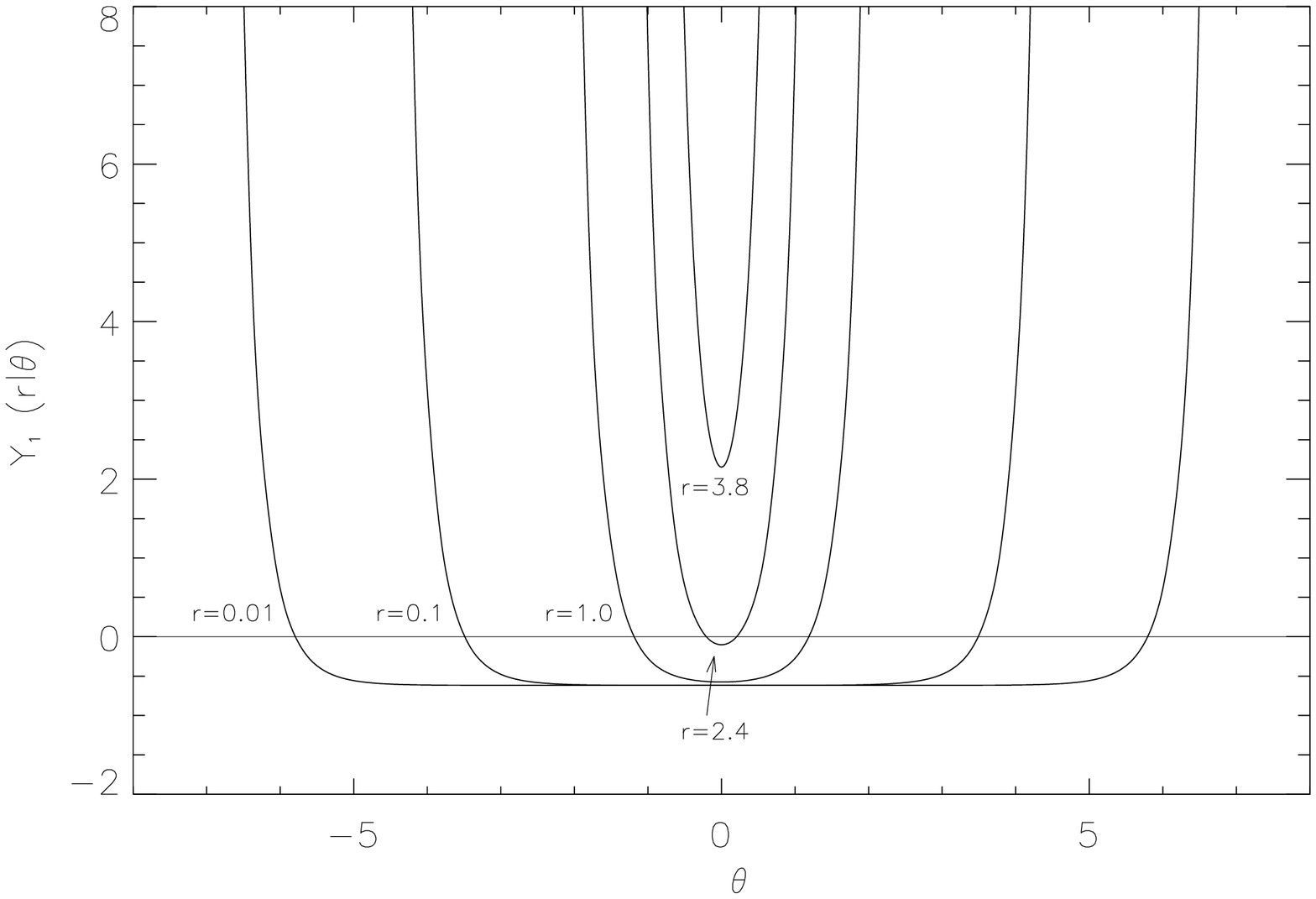}}
\capt{Fig.2. The function $Y_1(r|\theta)$ determined from the
numerical solution of the eqations \tbaee, \BAexcaa\ (for $r>r_0$) and 
(5.30), (5.31)\ (for $r<r_0$). }
\endinsert
\midinsert
\centerline{
\noindent\vbox{\offinterlineskip
\hrule
\halign{\vrule#&
  \strut\quad#\hfil\quad& 
        \vrule#& 
  \strut\quad#\hfil\quad& 
        \vrule#& 
  \strut\quad#\hfil\quad& 
        \vrule#& 
  \strut\quad#\hfil\quad&
        \vrule#\cr
height4pt&\omit&&\omit&&\omit&&\omit&\cr
&\hfil $r$ && $\hfil\gamma(r)$&&$\Delta E(r)/m$ && (TCSA) &\cr
height4pt&\omit&&\omit&&\omit&&\omit&\cr
\noalign{\hrule}
&$ 2.600000 $&&$ 0.133055 $&&$ 0.238050 $&&$ 0.23805 $&\cr
&$ 2.800000 $&&$ 0.256131 $&&$ 0.197403 $&&$ 0.19741 $&\cr
&$ 3.000000 $&&$ 0.322401 $&&$ 0.164321 $&&$ 0.16432 $&\cr
&$ 3.200000 $&&$ 0.367132 $&&$ 0.137217 $&&$ 0.13722 $&\cr
&$ 3.400000 $&&$ 0.399638 $&&$ 0.114891 $&&$ 0.11489 $&\cr
&$ 3.600000 $&&$ 0.424188 $&&$ 0.096412 $&&$ 0.096411 $&\cr
&$ 3.800000 $&&$ 0.443176 $&&$ 0.081057 $&&$ 0.081055 $&\cr
&$ 4.000000 $&&$ 0.458106 $&&$ 0.068256 $&&$ 0.068251 $&\cr
&$ 4.200000 $&&$ 0.469989 $&&$ 0.057552 $&&$ 0.057552 $&\cr
&$ 4.400000 $&&$ 0.479533 $&&$ 0.048579 $&&$ 0.048583 $&\cr
&$ 4.600000 $&&$ 0.487254 $&&$ 0.041042 $&&$ 0.041040 $&\cr
&$ 4.800000 $&&$ 0.493538 $&&$ 0.034700 $&&$ 0.034699 $&\cr
&$ 5.000000 $&&$ 0.498676 $&&$ 0.029356 $&&$ 0.029352 $&\cr
&$ 5.200000 $&&$ 0.502895 $&&$ 0.024846 $&&$ 0.024847 $&\cr
&$ 5.400000 $&&$ 0.506370 $&&$ 0.021037 $&&$ 0.021034 $&\cr
&$ 5.600000 $&&$ 0.509241 $&&$ 0.017816 $&&$ 0.017812 $&\cr
&$ 5.800000 $&&$ 0.511619 $&&$ 0.015092 $&&$ 0.015091 $&\cr
&$ 6.000000 $&&$ 0.513592 $&&$ 0.012785 $&&$ 0.012782 $&\cr
&$ 6.200000 $&&$ 0.515233 $&&$ 0.010832 $&&$ 0.010837 $&\cr
&$ 6.400000 $&&$ 0.516598 $&&$ 0.009178 $&&$ 0.009177 $&\cr
&$ 6.600000 $&&$ 0.517737 $&&$ 0.007775 $&&$ 0.0077767 $&\cr
&$ 6.800000 $&&$ 0.518688 $&&$ 0.006587 $&&$ 0.0065876 $&\cr
&$ 7.000000 $&&$ 0.519482 $&&$ 0.005581 $&&$ 0.0055816 $&\cr
&$ 7.200000 $&&$ 0.520147 $&&$ 0.004727 $&&$ 0.0047264 $&\cr
&$ 7.400000 $&&$ 0.520703 $&&$ 0.004003 $&&$ 0.0040045 $&\cr
&$ 7.600000 $&&$ 0.521169 $&&$ 0.003389 $&&$ 0.0033970 $&\cr
&$ 7.800000 $&&$ 0.521559 $&&$ 0.002870 $&&$ 0.0028753 $&\cr
&$ 8.000000 $&&$ 0.521886 $&&$ 0.002429 $&&$ 0.0024364 $&\cr
}
\hrule}
}

\capt{Table 5. The data obtained from the
numerical solution of\  \tbaee\  and \BAexcaa. The third column contains 
the excess energy  $\Delta E_1(r)$ defined by (5.36)
(given in units 
of the mass $m$) while the forth column contains the TCSA data for the same 
quantity computed in \Yur.}

\endinsert
As $r$ decreases the zeroes \eozero\  get closer together and at 
\eqn\rzero{r = r_0 \simeq 2.53576\ldots}
they collide on the real $\theta$-axis. For 
$r < r_0$ the function $Y_1 (r|\theta)$ has two real zeroes
\eqn\thetaoo{
\theta = \pm\ \alpha(r)\ ,}
and one  can represent it in the form:
\eqn\lmnbg{Y_1(r|\theta)=\sigma_0\big(\theta-\alpha(r)\big) \,
\sigma_0\big(\theta+\alpha(r)\big)
\ e^{\epsilon'_1(r|\theta)}\, ,\ \ \ \ \  r<r_0\  ,}
where the function $\sigma_0(\theta)$\ is
given by\  \sigmas.
Now we should solve  the equation
\eqn\TBAeA{\epsilon'_1(r|\theta) = r\cosh(\theta) +
\int_{-\infty}^{\infty}d\theta'\ \Phi(\theta-\theta')\log
\big(\sigma_0 (\theta'-\alpha(r))
\, \sigma_0 (\theta'+\alpha(r) ) +
e^{-\epsilon'_1(r|\theta')} \big)\ ,}
together with
\eqn\BAexactor{
i\tanh\big(
{{3\alpha(r)}\over 2}+i{\pi\over 4}\big)\,
e^{\epsilon'_1 (\alpha (r) +
i\pi/3)} = 
-i\tanh\big({{3\alpha(r)}\over 2}-i{\pi\over 4}
\big)\,e^{\epsilon'_1 (\alpha
(r) -i\pi/3)} = -1\ , }
instead of \tbaee\  and \BAexcaa.

\midinsert
\centerline{
\noindent\vbox{\offinterlineskip
\hrule
\halign{\vrule#&
  \strut\quad#\hfil\quad& 
        \vrule#& 
  \strut\quad#\hfil\quad& 
        \vrule#& 
  \strut\quad#\hfil\quad& 
        \vrule#& 
  \strut\quad#\hfil\quad& 
        \vrule#& 
  \strut\quad#\hfil\quad&
        \vrule#\cr
height4pt&\omit&&\omit&&\omit&&\omit&&\omit&\cr
&\hfil $r$ && $\hfil\alpha(r)$&&$\Delta E(r)/m$ && (TCSA) && (CPT) &\cr
height4pt&\omit&&\omit&&\omit&&\omit&&\omit&\cr
\noalign{\hrule}
&$ 1.0\,10^{-4} $&&$ 10.39926 $&&$ 23037.48 $&&$ \hfill $&&$ 23037.348 $&\cr
&$ 1.0\,10^{-3} $&&$ 8.096681 $&&$ 2302.837 $&&$ \hfill $&&$ 2302.834 $&\cr
&$ 1.0\,10^{-2} $&&$ 5.794092 $&&$ 229.3856 $&&$ \hfill $&&$ 229.3849 $&\cr
&$ 0.100 $&&$ 3.49150 $&&$ 22.0527 $&&$ \hfill $&&$ 22.0527 $&\cr
&$ 0.500 $&&$ 1.88203 $&&$ 3.67974 $&&$ \hfill $&&$ 3.67977 $&\cr
&$ 1.000 $&&$ 1.18800 $&&$ 1.44738 $&&$ \hfill $&&$ 1.44736 $&\cr
&$ 2.000 $&&$ 0.464431 $&&$ 0.430871 $&&$ 0.43088 $&&$ 0.430197 $&\cr
&$ 2.400 $&&$ 0.206153 $&&$ 0.288398 $&&$ 0.28839 $&&$ 0.286762 $&\cr
&$ 2.500 $&&$ 0.103278 $&&$ 0.261863 $&&$ *0.26189 $&&$ 0.259921 $&\cr
&$ 2.510 $&&$ 0.0875353 $&&$ 0.259338 $&&$ *0.25939 $&&$ 0.257393 $&\cr
&$ 2.520 $&&$ 0.0684264 $&&$ 0.256826 $&&$ *0.25691 $&&$ 0.254891 $&\cr
&$ 2.525 $&&$ 0.0565558 $&&$ 0.255570 $&&$ *0.25569 $&&$ 0.253650 $&\cr
&$ 2.530 $&&$ 0.0414288 $&&$ 0.254311 $&&$ *0.25447 $&&$ 0.252416 $&\cr
&$ 2.535 $&&$ 0.0150627 $&&$ 0.253036 $&&$ *0.25325 $&&$ 0.251188 $&\cr
}
\hrule}
}

\capt{Table 6. The data obtained from the
numerical solution of \TBAeA, \BAexactor. The third column contains 
the excess energy  $\Delta E_1(r)$ defined by (5.36)  (given in units 
of the mass $m$) while the forth column contains avaliable 
TCSA data for the same 
quantity computed in \Yur (the numbers marked with an asterisk obtained 
by the qubic spline interpolation of the resutls of \Yur). The fifth 
column contains CPT results  for the same quantity obtained from (5.37).}

\endinsert

The function $\alpha(r)$ obtained
by numerical integration of these equations is given in Table 6 for 
$10^{-4} \le r < r_0$. Note, that the  location of zeroes of $Y_1(r|\theta)$ 
for $r\simeq r_0$ both for $r>r_0$ and $r<r_0$ is described by the
same law (see Fig.3)
\eqn\zeroloc{
\left.{\alpha^2(r)\atop-\gamma^2(r)}\right\}\sim A\, (r_0-r), \qquad
r\simeq r_0,}
with the numerical value of constant $A=0.297893\ldots$.
\topinsert

\centerline{\epsfbox{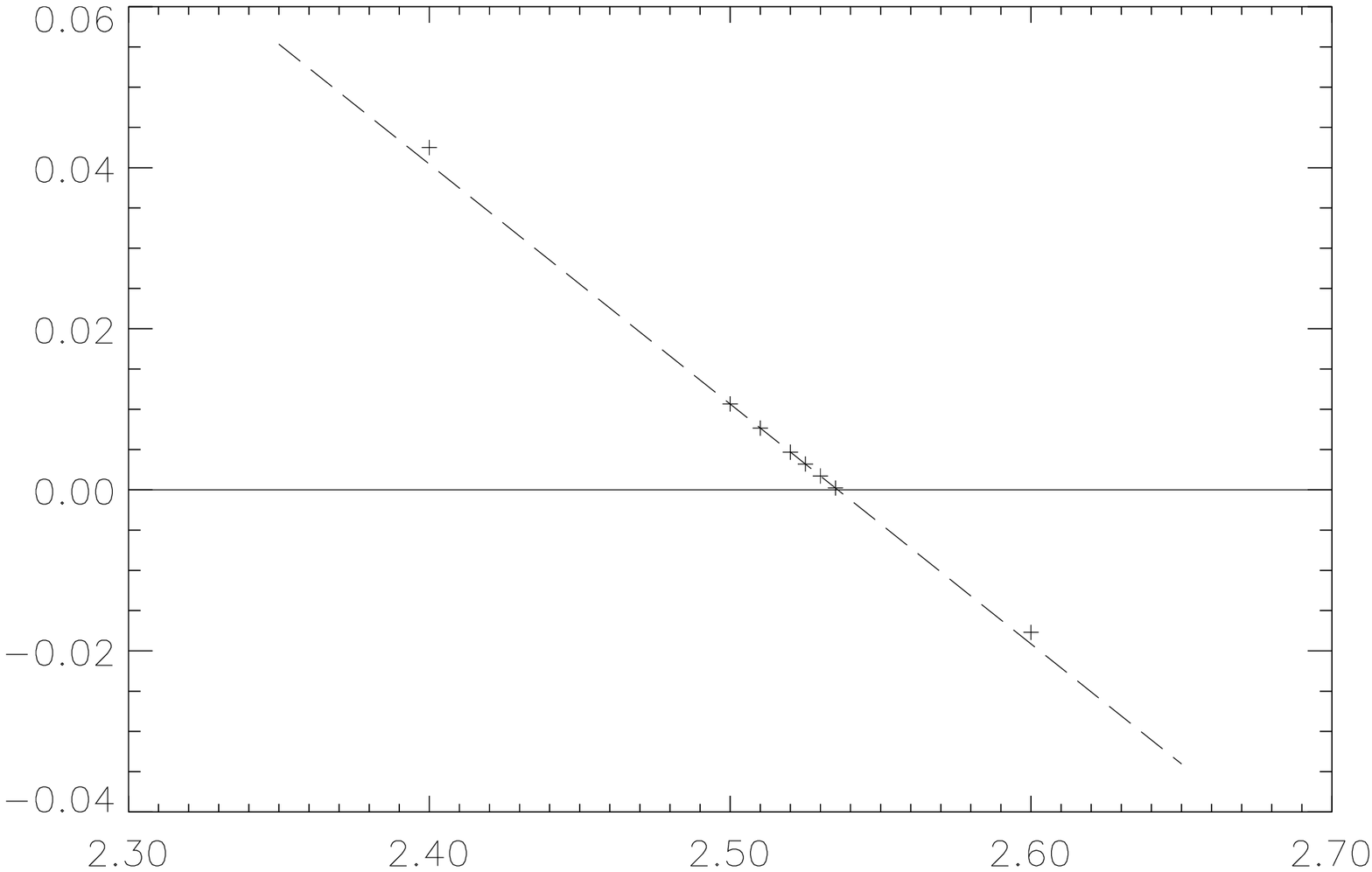}}

\capt{Fig.3. The values of $\alpha^2(r)$ and $-\gamma^2(r)$ for
$r\simeq r_0$ and the linear fit \zeroloc\ (dashed line).}
\endinsert

For $r \to 0$ the zeroes \thetaoo\  depart to 
$\pm\infty$ as
\eqn\depart{
\alpha(r) \sim -\log\big(\, {r\over 2\kappa}\,) + \alpha \, ,
\ \ r\to 0\ ,}
where the constant $\alpha$ is the same 
as in \thetaa. In this limit the function
$Y_1(r|\theta)$ approaches the constant $2\cos(2\pi p_1) =
(1-\sqrt{5})/2$ in the wide range of $\theta$,\  $\log r < \theta < -\log
r$, while  $Y_1(r|\theta) \sim Y_1\big(\theta +
\log\big( {r\over 2\kappa}\big)\big)$
for $\theta
\gtrsim -\log r$ and
$Y_1(r|\theta) \sim Y_1\big(-\theta 
-\log\big( {r\over 2\kappa}\big)\big)$
for $\theta \lesssim \log r$.
Here $Y_1(\theta)$ is the ``kink
solution'' discussed in Sect.4.
A typical form of the function
$Y_1 (r|\theta)$ for $r<r_0$ is shown in Fig.2.

The eigenvalues of the local IM ${\Bbb I}_{2n-1}$ associated with the 
excited state  $\mid \Psi_1\rangle$ can be calculated from the
asymptotic  expansion of $\epsilon_1 (r|\theta)$
for $r>r_0$ and  $\epsilon'_1 (r|\theta)$
for $r<r_0$  similar to \gdfd.
In particular, for the energy $ E_1(R)$,
$${\Bbb H} \mid \Psi_1\rangle=E_1 (R)\mid \Psi_1\rangle\ ,$$
this gives
\eqn\eofull{\eqalign{&
E_1 (R) = -{\sqrt{3}\over 12} \  m r +
2\,m\,\sin\big({\pi\over 6}+\nu (r)\big)
- {m\over 2\pi}\ 
\int_{-\infty}^{\infty}d\theta\ 
\cosh\theta\,\log(1+Y^{-1}_1 (r|\theta)), \ 
r>r_0\ ,\cr
&E_1 (R) = -{\sqrt{3}\over 12} \  m r +
\sqrt3 \, m\,\cosh\alpha(r)
- {m\over 2\pi}\
\int_{-\infty}^{\infty}d\theta\
\cosh\theta\ \log|1+Y^{-1}_1 (r|\theta)|,\ 
r<r_0\ ,}}

\topinsert

\centerline{\epsfbox{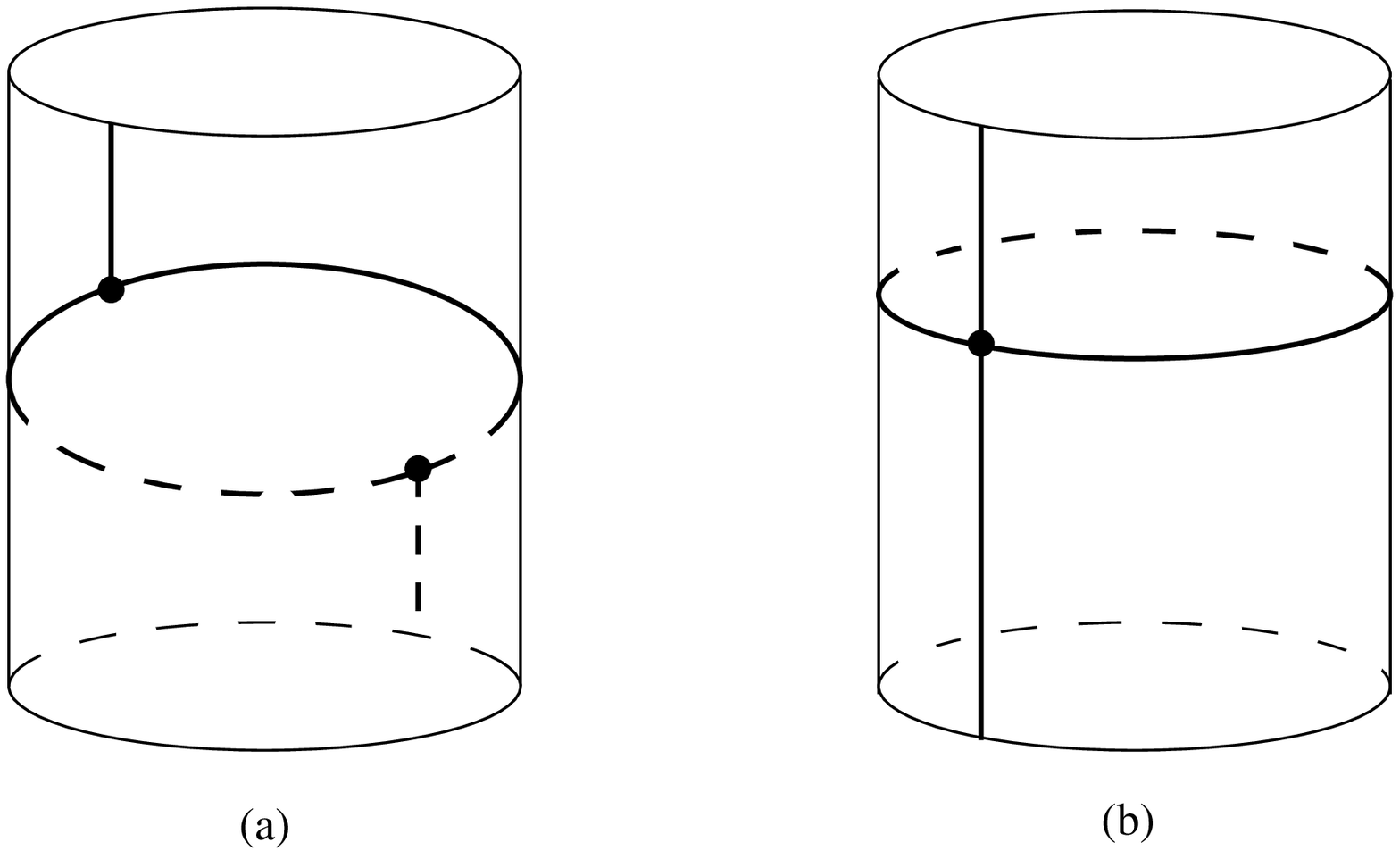}}
\capt{Fig.4. Diagramms representing two leading contributions to the large 
$r$ asymptotics of $\Delta E_1(r)$. Diagramms (a) and (b) coreresponds 
to the third and fourth terms in (5.35), respectively.}
\endinsert
\noindent Substituting \yoass\ 
and \nuass\ into the first of the above formulae
 one obtains a few first terms of
the large $r=m R$ expansion,
\eqn\eoexp{\eqalign{E_1 (R) =-{\sqrt{3}\over 12} \  m r& 
+ m + 3\, m\, e^{-{\sqrt{3}\over
2}\, r} -\cr
& {m\over 2\pi}\  \int_{-\infty}^{\infty}d\theta\ 
\cosh\theta\  S(\theta+i\pi/2)\ 
e^{-r\cosh\theta}
+ O\big(r e^{-\sqrt{3}\,r}\big)\ ,}}
where $S(\theta)$ is given by \Smatrix. Here the first term is the
bulk vacuum energy, the second term makes it manifest that for $R\to
\infty$ the state $\mid \Psi_1 \rangle$ describes one particle with zero
momentum, while the third and the fourth terms are the corrections
coming from the diagrams in Fig.4. 
These terms in \eoexp\  are in full 
agreement with the results of \ \Zar, \Yur\ and \KlassM. 

\topinsert

\centerline{\epsfbox{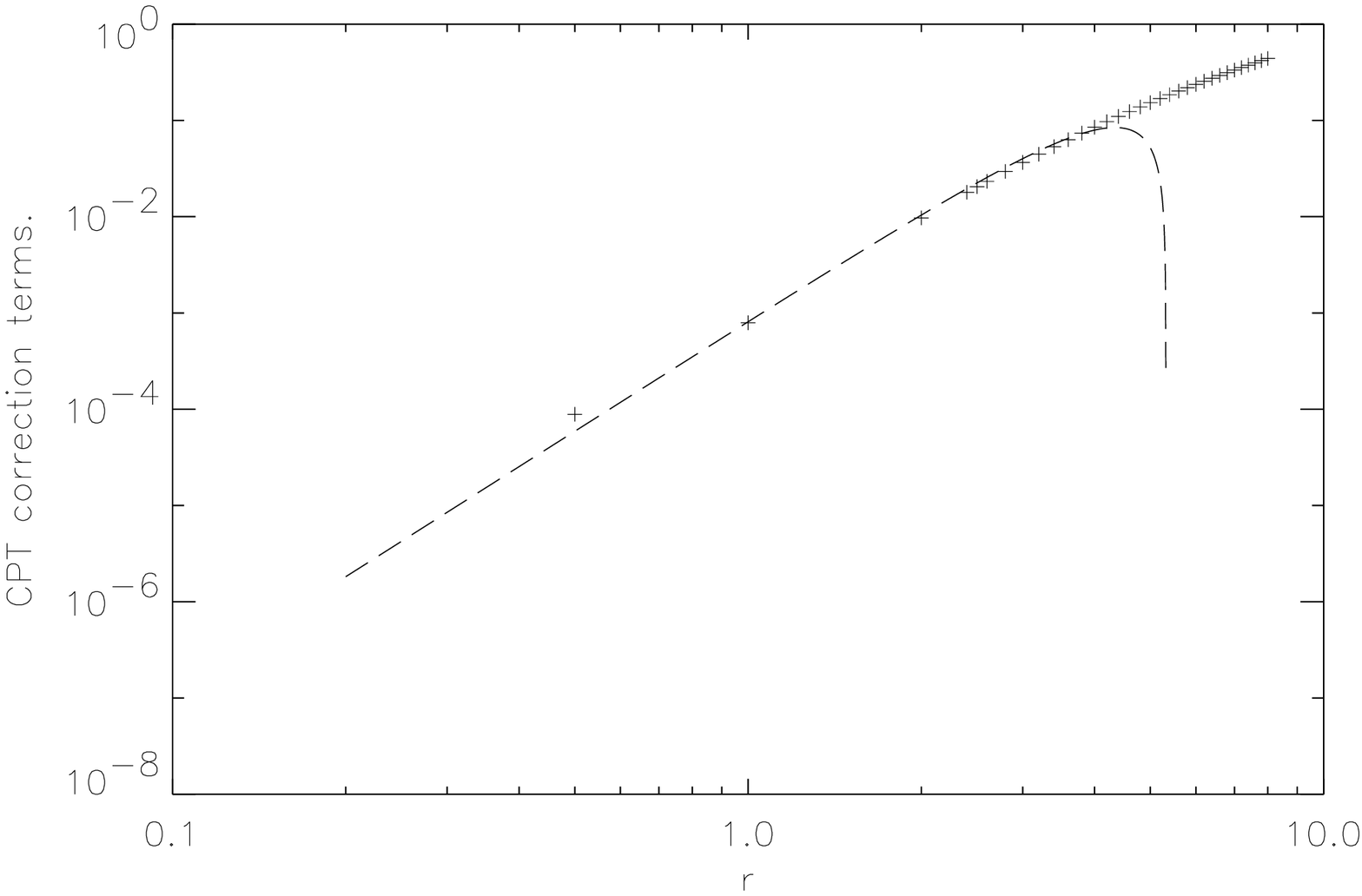}}
\capt{Fig.5. The dashed line represents the CPT correction terms 
(the second and third terms in (5.37) with negated signs given in
units of the mass $m$) and the points represent
the values for $11\pi/(15 r)-E_1(R)/m$ obtained from numerical
integration in \eofull.}
\endinsert
In the opposite limit $R\to 0$
the state 
$\mid \Psi_1 \rangle$ approaches the CFT highest weight state 
$\mid \Delta(p_1) \rangle $\ \erw\
in \hly. 
The later
has the dimensions $(\Delta, {\bar \Delta})
=(0,0)$ and corresponds to the identity operator in this CFT.
Indeed the calculation in Appendix C shows that
$E_1 (R) \sim 11\pi / (15\,
R), \ {\rm as} \ R\to 0.$
In Tables~5
 and 6  the function $E_1 (R)$ obtained by numerical integration in
\eofull\  is compared with the first excited state energy obtained in
\ \YY\ 
by Truncated Conformal Space Method. To facilitate this comparison 
we present the data for an ``excess energy'' 
\eqn\deltaE{\Delta E_1(r)=E_1(R)+{\sqrt{3}\over12} m r-m,}
where the bulk and mass terms are subtracted.
Also, in Table 6 we compare our
numerical results with the first terms of the short-distance expansion
\eqn\eoshort{E_1 (R) =  m\,\big( \ {{11\pi}\over {15\,r}} + 
e_2\,r^{19\over 5} + e_3\, r^{31\over 5} + \cdots\,  \big)\ ;}
$$
e_2 = -(1+\sqrt{5})^{3\over 2}\,
{5^{17\over 4}\over {2^{14}}}\,{{2^{1\over 10}}\over{\pi^{36\over5}}}\,
{{\Gamma^4 (2/5)}\over{\Gamma^2 (1/5)}}\,
\big(\Gamma(2/3)\,\Gamma(5/6)\big)^{24\over 5};
$$
$$
e_3 = 3\,(\sqrt{5}-1)^{1\over 2}\,{5^{29\over 4}\over 2^{25}}\,
{2^{9\over 10}\over \pi^{74\over 5}}\,\Gamma^4 (2/5)\,\Gamma^6 (1/5)\,
{{\Gamma^3 (9/10)\Gamma(3/10)}\over{\Gamma^3 (1/10)\Gamma(7/10)}}\,
\big(\Gamma(2/3)\Gamma(5/6)\big)^{36\over 5};
$$
obtained by Conformal Perturbation Theory from \SLYM\  (see Appendix  D).
In Fig.5 we plot numerical results for $E_1(R)/m-11\pi/(15 r)$ 
against the correction terms in \eoshort\foot{Note, that for $r\leq0.1$  
the correction terms in \eoshort\ are quite small
(less than $10^{-8}$ of the magnitude of the leading $1/r$ term) and
are beyond the precision of our numerical calculation.}.

\newsec{Discussion}

In Sect.3 we have constructed the commuting family of ${\Bbb
T}$-operators for perturbed CFT \action. As is well known the 
IQFT can be obtained by so called ``quantum group restriction'' of
the sine-Gordon model \ReSm\ \BeLeC\ . The sine-Gordon model is 
described by the action
\eqn\sg{{\cal A}_{SG} = \int\bigg[{1\over {16\pi}}\big(\partial_a \varphi
\big)^2 - 2 {\widetilde\mu}\, \cos(\beta\varphi)\bigg]\,d^2 x \ ,}
where $\varphi (z,{\bar z})$
is a scalar field and $\beta$ and ${\widetilde\mu}$
are parameters, the latter one carrying the dimension \foot{
In fact the value and even the dimension of the parameter ${\tilde \mu}$
depends on the precise way one defines the composite
field $\cos(\beta\varphi)$ in \sg. Here we assume that this field is
canonocally normalized with respect to its short-distance asymptotic,
i.e. $\cos(\beta\varphi)(z,{\bar z})\,\cos(\beta\varphi)(z',{\bar z}')
\sim {1\over 2}\,|z-z'|^{-4\beta^2}\, , \ 
(z,{\bar z})\to (z',{\bar z'})$. In this normalization the field
$\cos(\beta\varphi)$ carries the dimension $[\, length\, 
]^{-2\beta^2}$ and
hence the $\tilde \mu$ in \sg\ has the dimension (6.2).}
\eqn\mutilde{{\widetilde\mu} \sim \big[\, length\, 
\big]^{2\beta^2-2}.}
This QFT is integrable, in particular it possesses infinite set of 
local IM of the form (3.8) where now $T_{2k+2},\ \Theta_{2k} $ and 
${\bar T}_{2k+2},\ {\bar \Theta}_{2k}$ are 
certain local fields of the sine-Gordon model (and ${\hat \mu}$ is
replaced by ${\tilde \mu}$). The on-shell solution of \sg\ contains 
two ``topologically charged'' particles --- soliton and antisoliton --- and 
a number of the neutral particles with the masses
\eqn\masses{m_j = 2\,M \, \sin (\pi j \xi), \quad j={1\over 2}, 1, ... <
{1\over {2\xi}}\ ,} 
where $\xi$ is defined in \xibeta\ and $M$ is the soliton mass which in
turn is related to the parameter $\widetilde\mu$ in \sg as \Zarn\ 
\eqn\muM{M= {2\over \sqrt{\pi}}\,{{\Gamma({\xi\over 2})}\over
{\Gamma({{1+\xi}\over 2})}}\,\bigg[\pi{\tilde \mu}\,
{{\Gamma({1\over{1+\xi}})}\over
{\Gamma({\xi\over{1+\xi}})}}\bigg]^{{1+\xi}
\over 2} . }
The factorizable S-matrix of the sine-Gordon model can be found in \ZaZa\ .

In the case of infinite space the model \sg\ exhibits the symmetry
with respect to the quantum 
group $U_{\tilde q} (SL(2))$ \ReSm\ \BeLeC\  where
\eqn\dualq{{\tilde q} = e^{{i\pi}\over \beta^2}\ .}
Namely, the soliton and the antisoliton transform as the two-dimensional
representation of this quantum group and the local IM \pertloim\  and the
S-matrix commute with the generators ${\tilde E}, {\tilde F}$ and 
${\tilde H}$ of the associated quantum algebra $U_{\tilde q} (sl(2))$.
Naturally, the Hilbert space ${\cal H}_{\infty}$ of the infinite-space 
sine-Gordon 
model containes the subspace ${\cal H}_{\infty}^{singlet}$ of the states 
anihilated by the above generators. The remarkable fact discovered in 
\ \ReSm\ ,
\ \BeLeC\ 
is that this subspace ${\cal H}_{\infty}^{singlet}$ can be interpreted as
complete space of states of certain local QFT. It turnes out that this
``restricted sine-Gordon model'' coinsides with the perturbed CFT
\action, the
parameter ${\hat\mu}$ in \action\ being related to the constant ${\tilde
\mu}$ in \sg\ as \Zarn\ 
\eqn\mumuone{{\widetilde\mu}^2 = {{(1-2\beta^2)(3\beta^2-1)}\over \pi}\,
\bigg[{{\Gamma^3 (\beta^2) \Gamma(1-3\beta^2)}\over {\Gamma^3
(1-\beta^2) \Gamma(3\beta^2)}}\bigg]^{1\over 2}\, {\hat \mu}^2 }
Although the action of the generators ${\tilde E}, {\tilde F}, {\tilde
H}$ can be defined also in appropriate sectors of the finite-size 
sine-Gordon model with the 
spatial coordinate compactified on
a circle, the quantum group symmetry does not survive this
compactification\foot{The generators ${\tilde E}, 
{\tilde F}, {\tilde H}$ can be defined only in particular ``good"
sectors of the space of states ${\cal H}_{R}$ of the finite-size system, 
with appropriately chosen ``twists" (i.e. the eigenvalues of the operators
$\int_0^R \partial_x \varphi$ and $\int_0^R \partial_y \varphi$. This is 
similar to the supersymmetric QFT where in the finite-size case the SUSY 
generators exist only in the Ramond sector. However unlike the SUSY theories
in general case of $\tilde q$ the generators of $U_{\tilde q}(sl(2))$ change
the ``twists" and do not invariantly act in the ``good" sectors.}. 
Nontheless the 
notion of singlet states still exists, and the subspace ${\cal
H}_{R}^{singlet}\in {\cal H}_{R}$ still admits an interpretation as the 
space of states of the perturbed CFT \action\ in finite size geometry
\eqn\HH{{\cal H}^{singlet}_R \simeq {\cal H}_{PCFT}\ .}
It is clear from the structure of the operators ${\Bbb T}_j
(\mu|\lambda)$ defined in \Ljdef, \Ljbar\ and \fullT\ that their action 
extends to
the full sine-Gordon space ${\cal H}_{R}$ where they also satisfy
the commutativity conditions \fullcomm\ and all further relations in Sect.3
including \fullTexp\  and \fullfusion\  hold in the full
space. In other words the 
operators ${\Bbb T}_j
(\mu |\lambda)$ can be considered as the QFT ``transfer-matrices''
of the full unrestricted sine-Gordon theory. From this point of view the 
fields $\phi(z)$ and ${\bar\phi}({\bar z})$ admit very simple
interpretation in terms of the sine-Gordon field, namely $\phi$ and
${\bar\phi}$ are just the values of the field $\varphi(z, {\bar z})$ 
on the ``left'' and ``right'' components of the light cone, respectively,
i.e.
\eqn\sgphi{\phi(z) = {\beta\over 2}\varphi(z,0), \qquad {\bar\phi}({\bar z})
= -{\beta\over 2}\varphi(0,R-{\bar z});}
here we find it more convenient to think in terms of the Minkowski
space-time, where 
\eqn\xt{z=x+t, \qquad {\bar z}= x-t,}
are the real light-cone coordinates. Then according to \fullT\ the operator 
${\Bbb T}_j (\mu|\lambda)$ is obtained by integrating the 
Lax flat connection associated with sine-Gordon model (see e.g. \FaTa\ )
around the 
space-time cilinder, the integration being done in two stages --- first one
integrates along the ``left'' light cone from $z=0$ to $z=R$ with 
${\bar z}$ kept equal to $0$, and then the integration along the
``right'' light cone from ${\bar z}=0$ to ${\bar z}=R$ is
performed,\foot{This choice of the integration
path makes calculations simpler. As the connection is flat one could
have chosen any other closed contour which wraps around the
cilinder.} as is shown in Fig.6. 
\midinsert

\centerline{\epsfbox{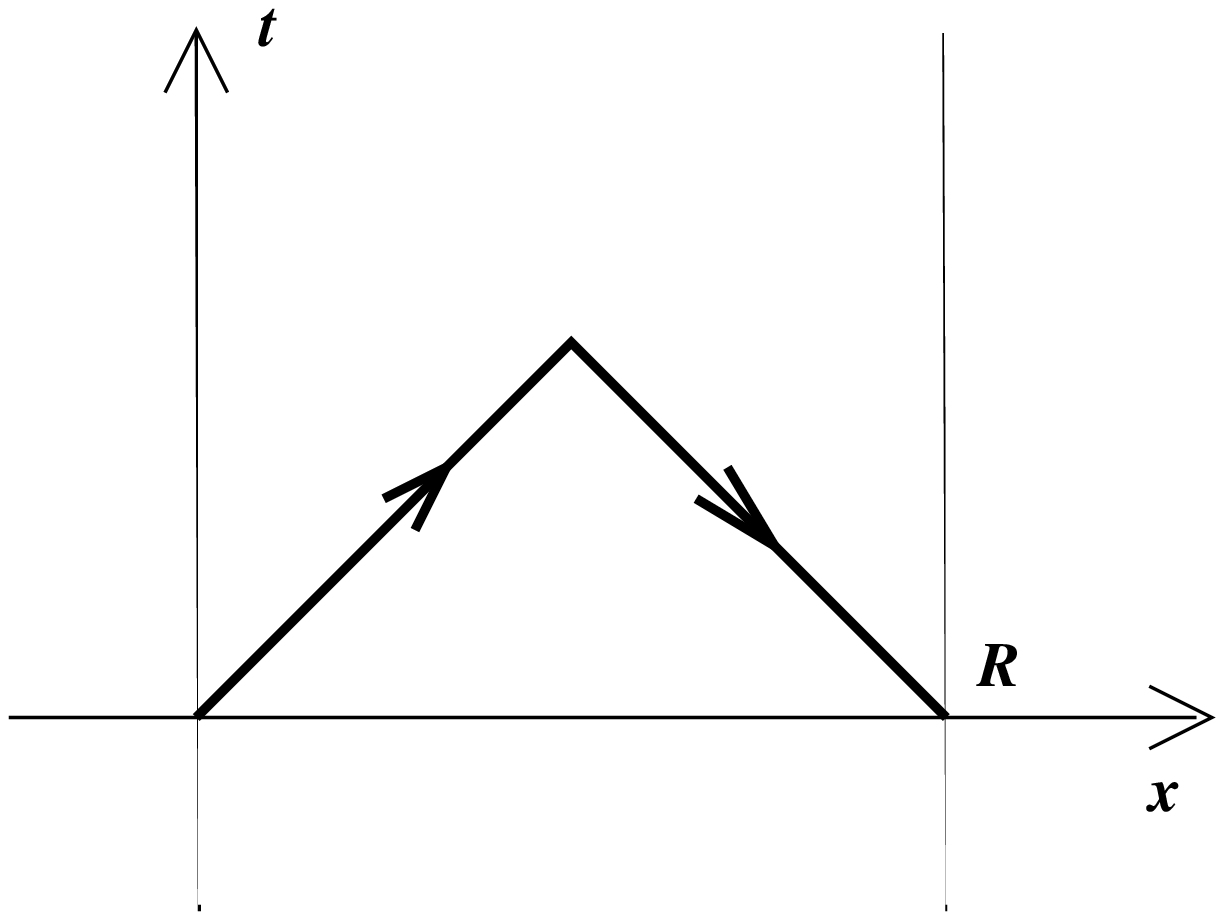}}

\capt{Fig.6. The integration path corresponding to the form \fullT\ of the 
sine-Gordon ``transfer matrix''.}
\endinsert

It is possible to check that if $\beta^2$ takes special values such that \qN\
is satisfied the truncation relation \Tred\ holds in the full sine-Gordon
space of states (the relation \Tredm\ is specific to the restricted
theory) and so an appropriate version of the method outlined in the Sectins
4 and 5 above can be applied to the sine-Gordon model at these values
of $\beta^2$. However for generic values of $\beta^2$ the approach based
on the QFT version of the Baxter's $Q$-operator is expected to be more
powerfull. 
We have shown in \BLZZ\ how to construct the ${\bf Q}$ operator in CFT, as
the trace similar to \Tjdef\ but this time taken not over a
finite-dimensional reppresentation of $U_q (sl(2))$ but over certain
infinite-dimensional representation of so-called $q$-oscillator algebra.
Clearly, the operator ${\Bbb Q}(\mu|\lambda)$ for the sine-Gordon
model can also be defined by the formula similar to \fullT. As we
observed in \BLZZ\ the asymptotic expansion of $\log {\bf Q}(\lambda)$ in 
CFT contains the contributions of the ``dual nonlocal IM'' in addition to
the local IM. It is easy to see that the ``dual nonlocal
IM'' are closely related to the nonlocal IM discovered in sine-Gordon
theory in \BeLeC\ . Therefore the operators ${\Bbb Q}(\mu|\lambda)$ and
associated Baxter's equations are expected to be a powerfull tools in
calculationg the spectra of both local and nonlocal IM in sine-Gordon
theory \foot{In fact, the relation \rel\ in our 
Conjecture 2 in Sect.3 is obtained
by combining \coeffcn, \muM\ and \mumuone.}. Our progress in this
direction is under way \BLZZZZ.

Our construction for the ${\Bbb T}$-operators can be extended to the
cases when higher rank quantum algebras replace $U_q (sl(2))$. In the case 
of CFT the operators ${\bf T}$ introduced in \BLZ\ was recently generalized
to the case of $q$-deformed twisted Kac-Moody algebra $A_{2}^{(2)}$ in \Rav. 
The ``massive''
versions of these operators similar to \fullT\ would apply to another class
of IQFT --- the CFT perturbed with the operator $\Phi_{1,2}$ --- which
includes in particular the $T=T_c$ Ising QFT with 
nonzero magnetic field \ABZ\ .

\vfill\eject
\centerline{\bf Acknowledgments}
\hskip0.5cm

The authors thanks B.D. 
Blackwell for many explanations on  computer graphics
software. 
S.L. acknowledges helpful discussions with A. LeClair.
The work of S.L. is supported in part by NSF grant.
A.Z. is grateful to the members of LPM, University of Montpellier,
for their kind hospitality which he enjoyed during the final stages 
of this work, and to Al. Zamolodchikov for sharing his insights. The 
research of A.Z. is supported by Guggenheim Fellowship and DOE grant
$\#$DE-FG05-90ER40559

\appendix{A}{}

In this Appendix we use the notation
\eqn\aa{a=\beta^2\ .}
While the simplest of the integrals \intgk\  $G_{1}^{vac} (p)$ is
easily evaluated,
\eqn\ab{\eqalign{G_{1}^{vac} (p)&= \int_{0}^{2\pi}\,dx\,
\int_{0}^{x}\,dy\,\bigg(\, 
2\,\sin\big(\,{{x-y}\over 2}\,\big)\, \bigg)^{-2a}\,2\,
\cos\big(2 p\,  (x-y-\pi)\big)\cr 
&=4\pi\,\sin\big(\pi(a-2p)\big)\
\int_{0}^{1}(1-t)^{-2a}\  t^{a-2p-1} \,dt \ =\ 
{{4\pi^2\, \Gamma(1-2a)}\over {\Gamma(1-a-2p)\, \Gamma(1-a+2p)}}\ ,}}
no explicit expression for these integrals with 
$k>1$ is known for generic $p$ and $a$. The integral $G_{2}^{vac}
(p)$ can be simplified as
\eqn\ac{\eqalign{
&G_{2}^{vac} (p) = (2\pi)^2\, \cos(2 \pi p)\  {\Gamma^2 (1-2a)\, 
\Gamma(a-2p)\,
\Gamma(a+2p)\over \Gamma(1-a-2p)\,\Gamma(1-a+2p)} \cr
&-2\pi\, \Gamma^2 (1-a)\,
\bigg(\,
{\sin\pi(a-2p)\ \Gamma(2p)
\over (a-2p)\ \Gamma(1-2a-2p)}\
F(a,p) +{ \sin\pi(a+2p)\ \Gamma(-2p)
\over (a+2p)\  \Gamma(1-2a+2p)}\
F(a,- p)\,\bigg)\ , }}  
where the function $F(a,p)$ is given by the series
\eqn\adas{F(a,p) = \Gamma(1+2a)\ 
\sum_{n=0}^{\infty}{{\Gamma(a-2p+n)}\over{n!\,
\Gamma(3a-2p+n)}}\  f_n (a,p)\ ,}
with
\eqn\adasaw{f_0 (a,p) = 1\, ,
\qquad f_1 (a,p) = 8a\ {{a-p}\over{1+a-2p}}\ ,}
and the rest of the coefficients
$f_n (a,p)$ with $n>1$ determined
recurrently through the relation
\eqn\hsyt{
A_n (a,p)\ f_{n+1} (a,p) - B_n (a,p)\ f_n (a,p) + C_n (a,p)\  f_{n-1}
(a,p) = 0\ ,}
where
$$A_n (a,p) = (1+n+a-2p)\,(1+n-4p);$$
$$B_n (a,p) = 2n^{3}+12n^2\,(a-p)+2n\,(8p^2+12a^2 -4ap
-7a)+8a\,(4p^2 + a -4ap -p);$$
$$C_n (a,p) = n\,(n+4a -1)\,(n+3a-2p-1)\,(n+4a-4p-1)\, .$$
This expression is very efficient for numerical evaluation of
$G_{2}^{vac}(p)$ with high precision.
Another convenient expression for
this integral which also generalizes to higher $G_{k}^{vac} (p)$ is
given in \ \SFN.

In particular case $2p=1-a$ the integrals $G_{2}^{vac}(p)$ and
$G_{3}^{vac}(p)$ can be evaluated analytically
\eqn\degst{G_{2}^{vac}\big({1-a\over 2}
\big) = {{2\pi^2}\over{1-2a}}\ {{\Gamma^3
(1-a)\, \Gamma(3-4a)}
\over{\Gamma^2 (2-2a)\, \Gamma(2-3a)}}\ ,}
\eqn\kdiu{G_{3}^{vac}\big({1-a\over 2}\big) =
{{4\pi^3}\over{1-2a}}\ {{\Gamma^3 (1-a)\,
\Gamma^2(1-2a)\, \Gamma({5\over 2}-3a)}
\over{\Gamma (2-4a)\, \Gamma(2-3a)\, \Gamma^3
({3\over 2}-a)}}\ .}

Finally, note that when $q$ is a root of unity the functional equations
\DNtba\ or \Mntba\ impose certain  
algebraic relations between the nonlocal IM, which
allow one to express some higher nonlocal IM through the lower ones. 
In particular, using the functional \YT\ one can show that 
\def\od{\textstyle{{1\over10}}}
\def\td{\textstyle{{3\over10}}}
\eqn\lyQ{\eqalign{
G^{vac}_2(\od)&={{5-\sqrt{5}}\over 10}\Big(G^{vac}_1(\od)\Big)^2\cr
G^{vac}_3(\od)& ={{\sqrt{5}-2}\over 5}\Big(G^{vac}_1(\od)\Big)^3\cr
}}
for the $p=1/10$, $\Delta=-1/5$ vaccuum state  in the Lee-Yang CFT and 
\eqn\lyQQ{\eqalign{
G^{vac}_4(\td)&={{5+\sqrt{5}}\over 10} \Big(
G^{vac}_2(\td)\Big)^2\cr
G^{vac}_5(\td)&={{5-\sqrt{5}}\over 10} \,
G^{vac}_2(\td)\,G^{vac}_3(\td)\cr
G^{vac}_6(\td)&={{5+\sqrt{5}}\over 10} \Big(
G^{vac}_3(\td)\Big)^2-\displaystyle
{{2+\sqrt{5}}\over 5}\Big(G^{vac}_2(\td)\Big)^3\cr
}}
for the $p=3/10$, $\Delta=0$ state.

\appendix {B}{}

Here we present  eigenvalues
of the
local IM  ${\bf I}_{2 n-1}$ ($n=1,2,...,8$)
with $R=2\pi$
on the
highest Virasoro vector with
the central charge $c$ and conformal dimension $\Delta$
\eqnn\ia
\eqnn\ib
\eqnn\ic
\eqnn\id
\eqnn\ie
\eqnn\if
\eqnn\ig
\eqnn\ih

$$\eqalignno{
I^{vac}_1(\Delta)&=\Delta-{c\over 24}\ ,&\ia\cr&&\cr
I^{vac}_3(\Delta)&=\Delta^2- {(c+2)\over 12}\ \Delta
+{c\, (5 c+22)\over 2880}\ ,&\ib\cr&&\cr
I^{vac}_5(\Delta)&=
\Delta^3-{(c+4)\over 8 }\ \Delta^2+
{(c+2)\, (3 c+20)\over 576}\
\Delta  -{c\, (3 c+14)\, (7 c+68)\over 290304
}\ ,&\ic\cr&&\cr
I^{vac}_7(\Delta)&=
\Delta^4-{(c+6)\over 6}\  \Delta^3+
{(15 c^2+194 c+568)\over 1440}\  \Delta^2-
{(c+2)\, (c+10)\, (3 c+28)\over 10368}\ 
\Delta&\cr
&\quad+{c\,  (3 c+46)\, (25 c^2+426 c+1400)\over 24883200}\ ,&\id\cr&&\cr
I^{vac}_9(\Delta)&=\Delta^5- {5\, (c+8)\over 24}\ 
          \Delta^4 +{(c+8)\, (5 c+46)\over 288}\  \Delta^3&\cr
&\quad-{(35 c^3+990 c^2+9048 c+23488)\over 48384}\  \Delta^2&\cr
&\quad+{(c+2)\, (175 c^3+7134 c^2+96168 c+392000)\over 11612160}\
         \Delta&\cr
&\quad      -{c\, (5 c+22)\, (11 c+232)\, 
        (7 c^2+274 c+1960)\over 3065610240}\ , &\ie\cr&&\cr
I^{vac}_{11}(\Delta)&=
   \Delta^6-{c+10\over 4}\  \Delta^5+
    {15 c^2+322 c+1808\over 576}\  \Delta^4&\cr
     &\quad-{105 c^3+3700 c^2+44612 c+165984\over 72576}\  \Delta^3&\cr
&\quad+{525 c^4+27908 c^3+548508 c^2+4248784 c+10147200\over 11612160}\
  \Delta^2&\cr
&\quad-{(c+2)\, (315 c^4+24604 c^3+676548 c^2+
7298480 c+25872000)\over 418037760}\ 
     \Delta&\cr
&\quad+{c\,  (13 c+350)\, (11025 c^4+1160780 c^3+25741404 c^2+
    198779728 c+470870400)\over 27389834035200}\ ,&\cr&&\if\cr}$$
\vfill
\eject
$$\eqalignno{I^{vac}_{13}(\Delta)&=
\Delta^7-{7\, (c+12)\over 24}\ \Delta^6+
{7\,  (15 c^2+ 386 c+2680)\over 2880}\ \Delta^5&\cr
&\quad-{(105 c^3 + 4430 c^2+ 66264 c+319552)\over 41472}\ \Delta^4&\cr
&\quad+ {( 525 c^4+ 33364 c^3+ 818172 c^2 + 8304848 c +27924096)
    \over 4976640}\ \Delta^3&\cr
&\quad-{(3465 c^5+331192 c^4+12012668 c^3+190959296 c^2+1291772608 c+
   2856307200)\over 1313832960}\  \Delta^2 &\cr
&\quad+(c+2)\, \big(121275 c^5+ 18838640 c^4+922934036 c^3 + 18803208352 c^2
         + 166162628800 c&\cr
&\phantom{\quad+(c+2)\, }\quad+517957440000\big)\ 
\Delta /3310859059200 &\cr 
&\quad-c\, ( 5 c+164)\,\big( 3465 c^5 + 1026934 c^4+39009476 c^3+
  568047656 c^2+ 3512182240 c&\cr
&\phantom{\quad-c\, ( 5 c+164)\, }\quad+7399392000\big)
/79460617420800\ ,
&\ig\cr&&\cr
I^{vac}_{15}(\Delta)&=
\Delta^8-{(c+14)\over 3}\,  \Delta^7+{7\, (c^2+30 c+248)\over 144}\ 
     \Delta^6&\cr
&\quad-{(7 c^3+344 c^2+ 6140 c+36304)\over 1728}\  \Delta^5&\cr
&\quad+{(35 c^4+ 2588 c^3+76020 c^2+952528 c+ 4102528)\over 165888}\ 
   \Delta^4&\cr
&\quad-{(231 c^5+ 25654 c^4+ 1122476 c^3+22259304 c^2+
     195500000 c+597940480)\over
     32845824}\   \Delta^3&\cr
&\quad+\big(21021 c^6+ 3835642 c^5+ 236079916 c^4+
     6438951928 c^3 + 84201249920 c^2&\cr
&\phantom{\quad+}\quad+
   500631533696 c+
1026155648000\big)\ \Delta^2/ 143470559232\ &\cr
&\quad -  (c+2)\big(3003 c^6 + 1137814 c^5 +
    89028148 c^4+2852262856 c^3+
    44100366464 c^2 &\cr
&\phantom{\quad -  (c+2)}
\quad+324626771840 c+ 897792896000\big)\ \Delta/ 1721646710784\ &\cr
&\quad+   {c\, (5 c+22)\, (7 c+68)\, 
   (17 c+658)\over 28097274319994880}\times&\cr
&\quad\qquad( 429 c^4+ 495724 c^3 + 20021388 c^2+
   285527760 c+1457456000)\ .&\ih\cr}$$

For  $c=-2$ the eigenvalues of
the all local IM is known \ \BLZZ:
\eqn\mcnb{I^{vac}_{2n-1}(\Delta)=2^{-n}\ B_{2 n}(2 p+1/2)\  ,\ \ \ 
{\rm with}\ \ \Delta=2 p^2-1/8\ ,}
where $B_k(x)$ is the  Bernoulli polynomials \Bern.

One can also find the eigenvalues\ $I^{vac}_{2 n-1}(\Delta)$\
for $c=1/2$ and $\Delta=0, 1/2, 1/16$:
\eqn\kdiu{\eqalign{&I^{vac}_{2n-1}(0)\big|_{c={1\over 2}}=
-c_n \ \big(1-2^{1-2n}\big)\ { B_{2 n}\over 4\, n}\ , \cr&
I^{vac}_{2n-1}(1/2)\big|_{c={1\over 2}}= c_n\  \Big(\, 2^{1-2 n}-
\big(1-2^{1-2n}\big)\ { B_{2 n}\over 4\, n}\, \Big)\ ,\cr
&I^{vac}_{2n-1}(1/16)\big|_{c={1\over 2}}=c_n\  {B_{2 n}\over 4\,  n}\ ,}}
Here $B_k$  is the  Bernoulli numbers\ \Bern\  and
$$c_n={(6 n-3)!!\  n!\  2^{n-1}\over
(2n-1)\,  (4 n-3)!\  3^n}\ .$$

\appendix {C}{}

In this Appendix we present the calculation of  first integrals 
in \kapaint\ and \kappaint\ (corresponding to the ground and the first 
excited state in the Lee-Yang CFT) using the so-called 
``dilogarithm trick''.
Although for the ground state case such calculations 
are well known in the literature on TBA (see e.g. \ 
\Sal) for the most 
transparent presentation) we include them here for a completeness.

Introducing the notation
\eqn\ldef{L_0(\t)=\log\big(1+e^{-\epsilon_0(\t)}\big)\ ,}
we can write the first integral in \kapaint\ as
\eqn\cala{\happa_1=\int_{-\infty}^{\infty}d\t\ e^{\t}L_0(\t)\ .}
Differentiating the integral equation \tba\ with respect
to $\t$, expressing  
therefrom the $e^{\t}$ term 
 and substituting the result into \cala\ one obtains 
\eqn\calb{ \happa_1\, \kappa= \int_{-\infty}^{\infty}d\t\  
\partial_{\theta}\epsilon_0(\t)\, L_0(\t)-
\int_{-\infty}^{\infty}d\t\int_{-\infty}^{\infty}d\t'\ 
\Phi(\t-\t')\  L_0(\t)\, \partial_{\theta'}L_0(\t')\ .}
Using now
the integral equation \tba\ to simplify the second term in the last 
formula one can rewrite it as 
\eqn\calc{\happa_1\, \kappa=  \int_{-\infty}^{\infty}
d\theta\ 
\big(\, \partial_{\theta}
\epsilon_0(\t)\, L_0(\t)-\epsilon_0(\t)\,\partial_{\theta}
L_0(\t)\, \big)
-\kappa\ \int_{-\infty}^{\infty}d\t\    e^{\t}L_0(\t)\ .}
Thus,
\eqn\cald{\happa_1\, \kappa= {1\over2}\ 
\int_{-\infty}^{\infty}d\t\ 
\big(\, 
\partial_{\theta}\epsilon_0(\t)\, L_0(\t)-
\epsilon_0(\t)\, \partial_{\theta}L_0(\t)\, \big)\ .}
A change to a new integration variable 
\eqn\vardef{
x(\t)=\big(1+e^{\epsilon_0(\t)}\big)^{-1}}
brings this this integral to the form
\eqn\cale{\happa_1\, \kappa=- 
{\rm Li\,}\big(x(\infty))+{\rm Li\,}(x(-\infty)\big) \ ,}
where ${\rm Li\,}(x)$ is the Rogers dilogarithm function
\eqn\rogers
{{\rm Li\,}(x)=-{1\over2}\int_0^x dy\  \Big(\, {\log(1-y)\over y}+
{\log(y)\over1- y}\, \Big)}
and 
$$
 x(-\infty)=  \Big({\sqrt{5}-1\over 2}\Big)^2\, , \qquad
x(\infty)=0\ ,
$$
as it follows from \yass\ and \yconstants. Here and below we will need
the following special values\foot{See, e.g.\  \Kirillov\  
for a comprehensive review on the
dilogarithm function.} of ${\rm Li\,}(x)$
\eqn\dilog{
{\rm Li\,}(1)={\pi^2\over 6},\qquad
{\rm Li\,}\Big({\sqrt{5}-1\over 2}\Big)={\pi^2\over 10},\qquad
{\rm Li\,}\Big(\Big({\sqrt{5}-1\over 2}\Big)^2\Big)={\pi^2\over 15}\ .}
Using this in \cale\ 
one arrives to the result \kapaa\ given in the main text.

Let us now turn to the calculation of the
first integral in \kappaint.
It can be conveniently rewritten as 
\eqn\calaa{{\tilde \happa}_1=
\int_{-\infty}^{\infty}d\t\ e^{\t}\, L_1(\t)\ ,}
where 
\eqn\lldef{L_1(\t)=\log\big(\sigma_0(\t-\alpha)+
e^{-\epsilon_1(\t)}\big)\ .}
with $\sigma_0(\theta)$ defined in \sigmas.
Proceeding as above we can
bring \calaa\ to the form similar to \cald\ 
\eqn\caldd{{\tilde \happa}_1\, \kappa= 
{{1}\over2}\int_{-\infty}^{\infty}d\t\ 
\big(\partial_{\theta}
\epsilon_1(\t)\, L_1(\t)-\epsilon_1(\t)\, \partial_{\t}
L_1(\t)\big)\ .}
Defining new functions 
\eqn\fundef{\eqalign{
\ep(\t)&=\epsilon_1(\t)+\log|\sigma_0(\t-\alpha)|=\log|Y_1(\t)|\, ,\cr
\ol(\t)&=L_1(\t)-\log|\sigma_0(\t-\alpha)|=\log|1+Y_1^{-1}(\t)|\ ,}}
one can further rewrite \caldd\ as
\eqn\calcc{{\tilde \happa_1}\, \kappa= I+
{1\over2}\int_{-\infty}^{\infty}d\t\ 
\Big(\partial_{\t}\ep(\t)\,\ol(\t)-
\ep(\t)\,\partial_{\t}\ol(\t)\Big)\ ,}
where $I$ is determined as the principal value of the singular integral
\eqn\dovesok{
I=-\  \vpint  d\t\  \big(L_1(\t)+\epsilon_1(\t)\big) \,
\partial_\t\log|\sigma_0(\t-\alpha)|\ .}
The integral in \calcc\ is calculated similarly to that in \cald.
Splitting the interval of integration into two parts, $(-\infty,\alpha)$
and  $(\alpha,\infty)$ and changing to a new integration variable 
\eqn\defvars{
\tilde{x}(\theta) =\cases{e^{\ep(\t)},&for $\t<\alpha$ \cr
\big(1+e^{\ep(\t)}\big)^{-1},&for $\t>\alpha$\ ,\cr}}
one can bring \calcc\ to the form
\eqn\calee{\tilde{\happa}_1\, \kappa= I+
{\rm Li\,}(1)+{\rm Li\,}\big(\,\tilde{x}(-\infty)\, \big)\ ,}
where ${\rm Li\,}(x)$ is the dilogarithm function \rogers\ and 
\eqn\xlim{
\tilde{x}(-\infty)={\sqrt{5}-1\over 2}}
as it follows from \yconstants. The remaining integral \dovesok\
is calculated by 
using the integral equation  
\tbaa\  and the condition \baaa\ 
determining the root 
$\alpha$.
In fact, using \tbaa\ first write \dovesok\ in the form
\eqn\calf{
I= - \pi\sqrt{3 }\, \kappa\, e^{\alpha}-\ 
\vpint \vpint d\t d\t'\ 
L_1(\t)\,\big(\Phi(\t-\t')+\delta(\t-\t')\big)\ 
\partial_{\t'} \log|\sigma_0(\t'-\alpha)|\ ,}
where we have used the elementary integral
\eqn\elemint{
\vpint 
d\t\  e^\t \partial_\theta 
\log|\sigma_0(\t-\alpha)| =\pi\sqrt{3}\, e^{\alpha}\ .}
The integral over $\t'$ in \calf\ is also elementary and 
can be easily calculated. After that the r.h.s. of \calf\ becomes 
(up to a simple factor) identical to the l.h.s. of \baaa\ 
\eqn\calg{
I= - \pi\sqrt{3}\,\kappa\, 
e^{\alpha}-3\ \vpint d\t\ {\cosh 2(\t-\alpha)\over
\sinh3(\t-\alpha) }\ L_1(\t)=-\pi^2\ (1+4N)\ .}
Collecting together \calee,\  \calg,\ \dilog\    one obtains 
\eqn\otvet{
{\tilde \happa}_1\, \kappa=
- \Big({22\over5}+24\, N\Big)\  {\pi^2\over 6}\ ,}
which for $N=0$ precisely gives \tkappaa .
The above calculations can easily be generalized for 
the solutions of \TBA\  and \BAdress\  with arbitrary number of (real and 
complex) zeroes. Similar calculations for in lattice models 
can be found in \ \KluP.

\appendix {D}{}

One obtains Conformal 
Perturbation Theory (CPT) by expanding in the interaction
parameter ${\hat \mu}^2$ in \action. The CPT 
for the ground-state energy $E_0 (R)$ of \SLYM\
was discussed in \ \Zar.
It can be easily modified for the excited-state energies. The excited
-state energy $E_1 (R)$ in Sect.5 expands as
\eqn\eoo{\eqalign{&E_1 (R) = -{\pi c\over 6R}-
R\,\sum_{k=2}^{\infty}{\big(-{\hat \mu}^2\big)^k\over k!}\
\Big(\, {2\pi\over
R}\, \Big)^{2k\, (\Delta -1)+2} \times\cr
&\int
\prod_{i=2}^{k}\big(\, 
(z_i {\bar z}_i)^{\Delta-1}\, d^2 z_i\, \big)\ 
 \langle \Phi (1,1) \Phi(z_2, {\bar z}_2 )\Phi(z_3, {\bar z}_3 )
\dots \Phi(z_k, {\bar z}_k ) 
\rangle_{CFT}^{conn}\ ,}}
where $d^2 z\equiv dx\,dy$\  for $z=x+iy$, \ 
$c=-22/5$ is the central charge of CFT ${\cal M}_{2/5}$, $\Delta
= -1/5$ is the conformal dimension of the field $\Phi$ in \ \SLYM\  and 
$\langle \dots \rangle_{CFT}^{conn}$ are the (connected) correlation 
functions of the this CFT \foot{For generic excited-state
energy $E_n (R)$ the expansion similar to \ \eoo\  would contain the 
insertion $V_n (0,0) V_n (\infty, \infty)$ in the correlation functions,
where $V_n (z, {\bar z})$ is the CFT fields associated with the CFT limit
of the excited
state $\mid \Psi_n \rangle$ under consideration. For the state
$\mid\Psi_1\rangle_{R=0} $ the associated field is the 
identity operator and therefore this insertion is ignored in \eoo.}. 
The first terms in \eoo\  contain
\eqn\eou{\langle \Phi(z_1, {\bar z}_1)\Phi(z_2, {\bar z}_2)
\rangle_{CFT}^{conn} = |z_1 - z_2|^{-4\Delta}\ ,}
\eqn\kiuy{\langle \Phi(z_1, {\bar z}_1)\Phi(z_2, {\bar z}_2)\Phi(z_3, {\bar
z}_3) \rangle_{CFT}^{conn} = {\Bbb C}_{\Phi \Phi \Phi}\,
|z_1 - z_2|^{-2\Delta}\, |z_2 - z_3|^{-2\Delta}\,|z_3 - z_1|^{-2\Delta}\ .}
Here
$${\Bbb C}_{\Phi \Phi \Phi} = i\, 
{\, 5^{1\over 4} \over {10\pi}}\
\Gamma^3 ({1/5})\Gamma({2/5})
$$
is the triple-$\Phi$ OPE structure constant \ 
\DotsFat. The most important 
difference in \eoo\  
as compared with the similar
formula for $E_0 (R)$ 
in \ \Zar\ 
is that the integrals in \eoo\  diverge as $(z_i, {\bar z}_i)\to 0,
\infty$ and therefore the expression \eoo\  can not be taken literally.
This divergence is easily traced down to the fact that $\mid \Psi_1
\rangle$ is not a ground state. As is well known in ordinary quantum 
mechanical perturbation theory, in doing calculations for the excited
states in each order one has to take care of the orthogonality between 
the state under consideration and all the lower-energy states. This
condition brings into \ \eoo\  additional terms which exactly provide
the subtractions of the above divergences. So in \eoo\  appropriate
subtractions of these divergences are implied. Equivalent way to
define the integrals in  \eoo\  (which is similar to ``dimensional
regularization'' of Feynman diagrams) is to evaluate the integrals in
\eoo\  for $\Delta <1/2$ where the integrals converge and then
analytically continue to $\Delta = -1/5$, see e.g. \Confor. 
With this prescription the
integrals in the first two terms of the series in \eoo\  can be evaluated
explicitly,
\eqn\tredf{\int d^2 z\ 
 |z|^{2\Delta - 2}\, |1-z|^{-4\Delta} = \pi
\ {{\Gamma^2 (\Delta)\, \Gamma(1-2\Delta)}\over {\Gamma^2 (1-\Delta)
\, \Gamma(2\Delta)}}\ ,}
\eqn\kjhgf{\eqalign{\int
d^2 z_1 \,d^2 z_2\ 
|z_1|^{2\Delta -2}\,& |z_2|^{2\Delta -2}\, |1-z_1|^{-2\Delta} 
\,|1-z_2|^{-2\Delta}\, |z_1 - z_2|^{-2\Delta}=\cr
&{\pi^2\over 4}\, {{\Gamma^3 ({\Delta\over 2})\, 
\Gamma (1-{{3\Delta}\over
2})}\ 
\over {\Gamma^3 (1 - {\Delta\over 2})\,
\Gamma ({{3\Delta}\over 2})}}\ }}
and one arrives at \eoshort.

\listrefs

\end